%% file: bare_jrnl.tex
\documentclass[journal]{IEEEtran}
\usepackage[usenames,dvipsnames,svgnames,table]{xcolor}
\usepackage{ulem}
\ifCLASSINFOpdf
  \usepackage[pdftex]{graphicx}
  \DeclareGraphicsExtensions{.pdf,.jpeg,.png}
\else
\fi
\usepackage{subfig}
\usepackage{epstopdf}

\usepackage[cmex10]{amsmath}

\usepackage{stfloats}
\usepackage{url}
\usepackage{multirow}

\usepackage{xspace}

\newcommand{\etal}[1]{#1~{et~al.}}

\begin{document}
%
\title{Humans Are Easily Fooled by Digital Images}
%
%
%

\author{Victor~Schetinger,
				Manuel~M.~Oliveira,
        Roberto~da~Silva,
				and~Tiago~Carvalho
				}

%
%


\markboth{Transactions on Information Forensics and Security}%
{Shell \MakeLowercase{\textit{et al.}}: Bare Demo of IEEEtran.cls for Journals}
%



\maketitle

\begin{abstract}
Digital images are ubiquitous in our modern lives, with uses ranging from social media to news, and even scientific papers. For this reason, it is crucial evaluate how accurate people are  when performing the task of identify doctored images. In this paper, we performed an extensive user study evaluating subjects capacity to detect fake images. After observing an image, users have been asked if it had been altered or not. If the user answered the image has been altered, he had to provide evidence in the form of a click on the image. We collected 17,208 individual answers from 383 users, using 177 images selected from public forensic databases. Different from other previously studies, our method propose different ways to avoid lucky guess when evaluating users answers. Our results indicate that people show inaccurate skills at differentiating between altered and non-altered images, with an accuracy of 58\%, and only identifying the modified images 46.5\% of the time. We also track user features such as age, answering time, confidence, providing deep analysis of how such variables influence on the users' performance.
\end{abstract}

\begin{IEEEkeywords}
Digital Image Forensics, Behavior Forensics
\end{IEEEkeywords}

%
\IEEEpeerreviewmaketitle

\input{sec/introduction}

\input{sec/methodology}
\input{sec/results}

\input{sec/relatedwork}
\input{sec/conclusion}

\input{sec/appendix2}



\ifCLASSOPTIONcaptionsoff
  \newpage
\fi



\bibliographystyle{IEEEtran}
%
%
%

\bibliography{bare_jrnl}

%








\end{document}

%% file: sec/introduction.tex
\section{Introduction}
\label{sec:Introduction}

In July 2010 the Australian newspaper \textit{Sydney Morning Herald} published  news about Dimitri De Angelis' case. According to this news source, Mr. De Angelis deceived 10 people into investing money on him, raising over 7 millions dollars in the course of 5 years. To convince investors, Mr. De Angelis presented himself as a successful music producer and, to prove he was an influential individual, he sent photographs of himself side by side with people as Pope John Paul II, Alan Greenspan, Bill Clinton and Bill Gates\footnote{http://www.smh.com.au/nsw/im-a-fun-guy-not-a-fraudster-20100703-zuz6.html - Accessed: Feb. 2015}. Unfortunately for his investors, the pictures were all digital forgeries. Mr. De Angelis was sentenced, in March 2013, to twelve years in prison\footnote{http://www.fourandsix.com/photo-tampering-history/?currentPage=18  - Accessed: Feb. 2015}.

This is just one of many examples showing how humans are easily fooled by digital image manipulations. Even a doctored image created as a joke has the potential to easily spread on social media and cause misinformation. The idea that people are not suited to assess an image's authenticity without the aid of additional tools is widely explored by the forensics community~\cite{SurveyPiva}~\cite{Rocha_2011}~\cite{FontaniDataFusion}.

However, there is little experimental research focused on identifying this lack of perception related to digital forgeries. Research on human perception of digital images focuses on very specific aspects of vision such as lighting, geometry, and face recognition, usually having a limited experimental scope. No deep study has been performed to explicitly evaluate people's skills in the task of finding forgery in digital images.

In this work we look for evidence to help the forensics community to understand how users identify forgeries in digital images, and how accurate they are. We constructed a test set, and performed a perception experiment with approximately 400 users. Not surprisingly, the average accuracy in determining if an image is fake or not is only slightly better than random chance, with users guessing right around 58\% of the time. To make the test as relevant as possible for the forensics community, our test was performed using images from known public forensics data sets and designed to gather as much input as possible.

Our study differs from previous ones because it asks users to provide evidence (in the form of a click) when they believe the image has been altered. This allows us not only to distinguish between lucky guesses and correct answers, but it also provides insight on what subjects perceive as suspect in an image. Furthermore, by implementing the study as an online test, we were able to gather more data than any previous work. Our main contributions can be summarized as:
%
\begin{itemize}
\item We present strong evidence that users have difficulty at identifying forgery in digital images, even in a context where they have been explicitly told to look for it (Section ~\ref{sec:Overview});
\item We show that an user's background, such as age and education, has a small effect on his performance, and how the answering behavior, such as timing and confidence, affects the result (Section ~\ref{sec:UserBackground});
\item We demonstrate that there is meaningful difference in the hit rate for the most common types of forgery in the forensics literature: erasing, copy-pasting, and splicing (Section ~\ref{sec:Images});
\item We provide a public data set of user answers for real and forged images\footnote{The dataset will be made available upon paper acceptance.}, which to the best of our knowledge is the of its kind available.
\end{itemize}

%% file: sec/methodology.tex
\section{The User Study}
\label{sec:Methodology}

The main objective of this study is to assess how hard it is for an average individual to determine when an image has been modified. 
For this, we gathered input from a large group of subjects about images from a large database.

Subjects are shown one image at a time and asked to provide a binary \textbf{yes}/\textbf{no} answer to the following question: "Is there any kind of forgery present in this image?". 
%
For simplicity, we will call an {\it authentic} image (also referred to as pristine or original, in the forensics literature) as a \textbf{T} ({\it true}) image.
Likewise, we will call a {\it modified} image (also denoted forged, tampered, fake or edited) as an  \textbf{F} ({\it false}) image.

If a subject answers \textbf{yes}, he is saying the image is {\it false}, and we call this an \textbf{F} answer, as opposed to a \textbf{T} answer. In this case, he is asked to provide evidence that the image has been altered. This evidence is provided in the form of a click on the image, pointing to a region that indicates it has been altered. Different forms of evidence are considered valid, such as the altered region itself, its close surroundings or even irregular shadows left by the forgery. For F images, the user is considered to have answered correctly only if valid evidence is provided.

Considering all the different answer combinations, there are five possible outcomes: the image can be either T or F, the user answer can be either T or F, and if the user answers F, he can either provide valid or invalid evidence. To refer to all the cases in a simple fashion, the notation "Image Type":"Answer Type" is used. This can be seen in detail in Table~\ref{tab:answerClasses}, and the different cases are each called \textbf{answer classes}. This format will also be used thorough the paper for presenting results.

\begin{table*}[ht]
\centering
    \begin{tabular}{|l|l|l|l|}
    \hline
    Class & Meaning                                                                   & Answer    & Type           \\ \hline
    T:T   & The image is T and the user provided a T answer.                     & Correct   & True Negative  \\
    F:Fv  & The image is F, the user provided a F answer and valid evidence.   & Correct   & True Positive  \\
    F:T   & The image is F and the user provided a T answer.                    & Incorrect & False Negative \\
    F:Fi  & The image is F, the user provided a F answer and invalid evidence. & Incorrect & False Negative \\
    T:F   & The image is T and the user an F answer.                       & Incorrect & False Positive \\ \hline
    \end{tabular}
		\caption{Different answer classes for the user study, in the notation "Image Type":"Answer Type". }
		\label{tab:answerClasses}
\end{table*}

\textbf{We treat the users' answers as a binary classification problem where they are trying to identify F images}. An image is correctly identified as F by providing an F answer with valid evidence (F:Fv), corresponding to a true positive. Determining an image is T and does not contain alteration is a true negative, denoted by the class T:T. A false positive is when an image is T, but an user classifies it as F (T:F), and a false negative is when he fails to properly identify an F image (F:T). Answering an F image is F, but providing invalid evidence (F:Fi) is considered to be a a false positive, because the actual altered region was not identified by the user. 

\subsection{Implementation}

The user study was implemented in the form of a website\footnote{http://newton.inf.ufrgs.br}. Users were asked to register, providing background information such as age, education and experience with digital images. Once registered, users could log in at any time to analyze and answer images, and log out to come back at later time. 

\begin{figure}[ht]
  \centering
  \includegraphics[width=0.48\textwidth]{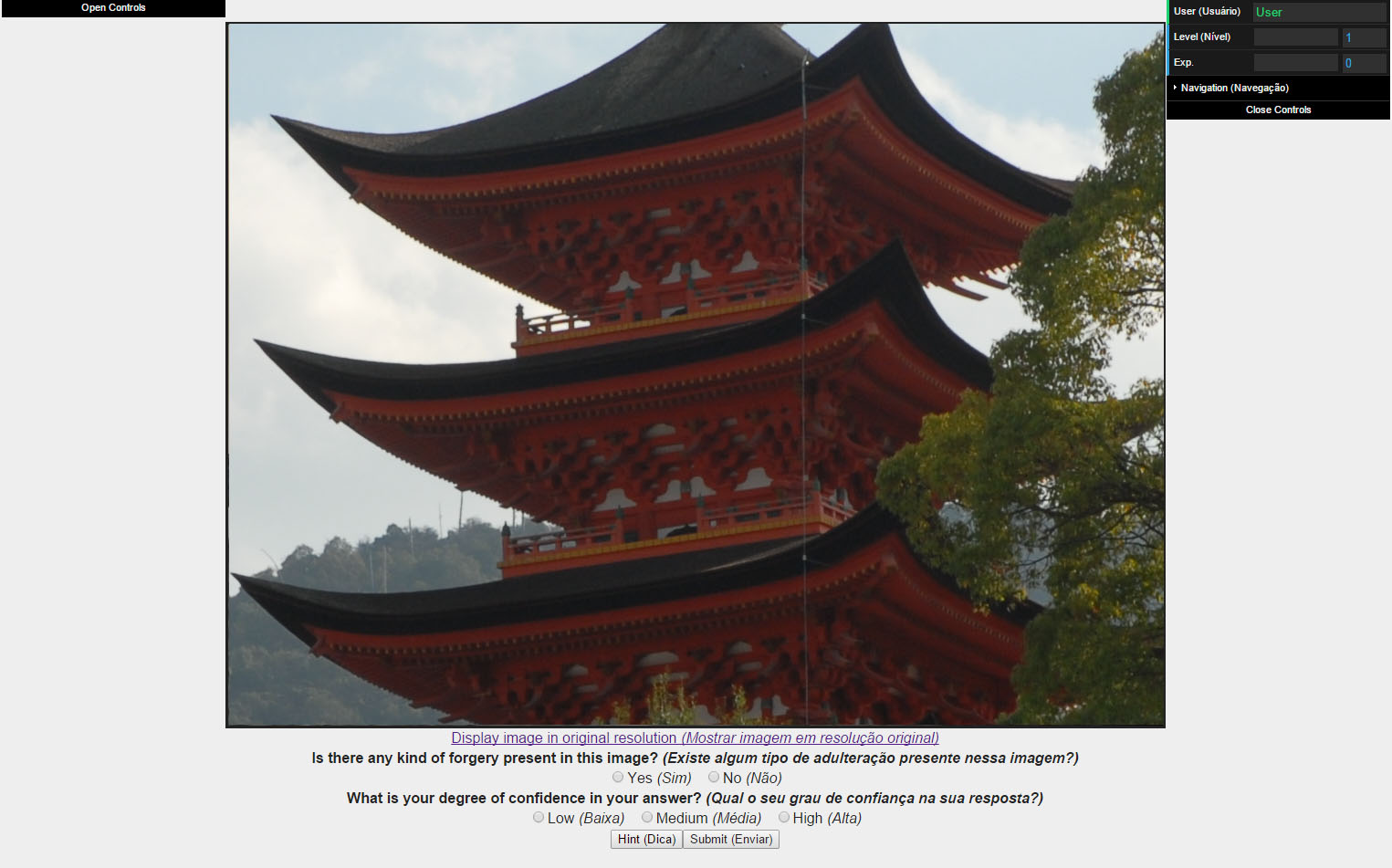}
  \caption{Interface for the online test.}
	\label{fig:SiteExample1}
\end{figure}

The answering form consisted of a simple web page, as depicted in Figure~\ref{fig:SiteExample1}. The current image being evaluated by the user is displayed at the page's center with the form to check the answer and user's confidence at the bottom. If the user is in doubt, after waiting for 20 seconds he is allowed to ask for a hint, removing a region of half the image area where there were no alterations present\footnote{In the case of real images, a random region of half the image area is discarded}. 

On the top right of the web page a menu displays the current user progress in the test. This menu can also be used to access other useful locations in the test website such as the tutorial and contact pages. On the top left, there is an advanced menu, which is minimized by default, that allows the user to provide additional insights on what he feels might be wrong with the image.

\begin{figure*}[ht]
\centering
\subfloat[Image without hint.]{\includegraphics[width=1.5in]{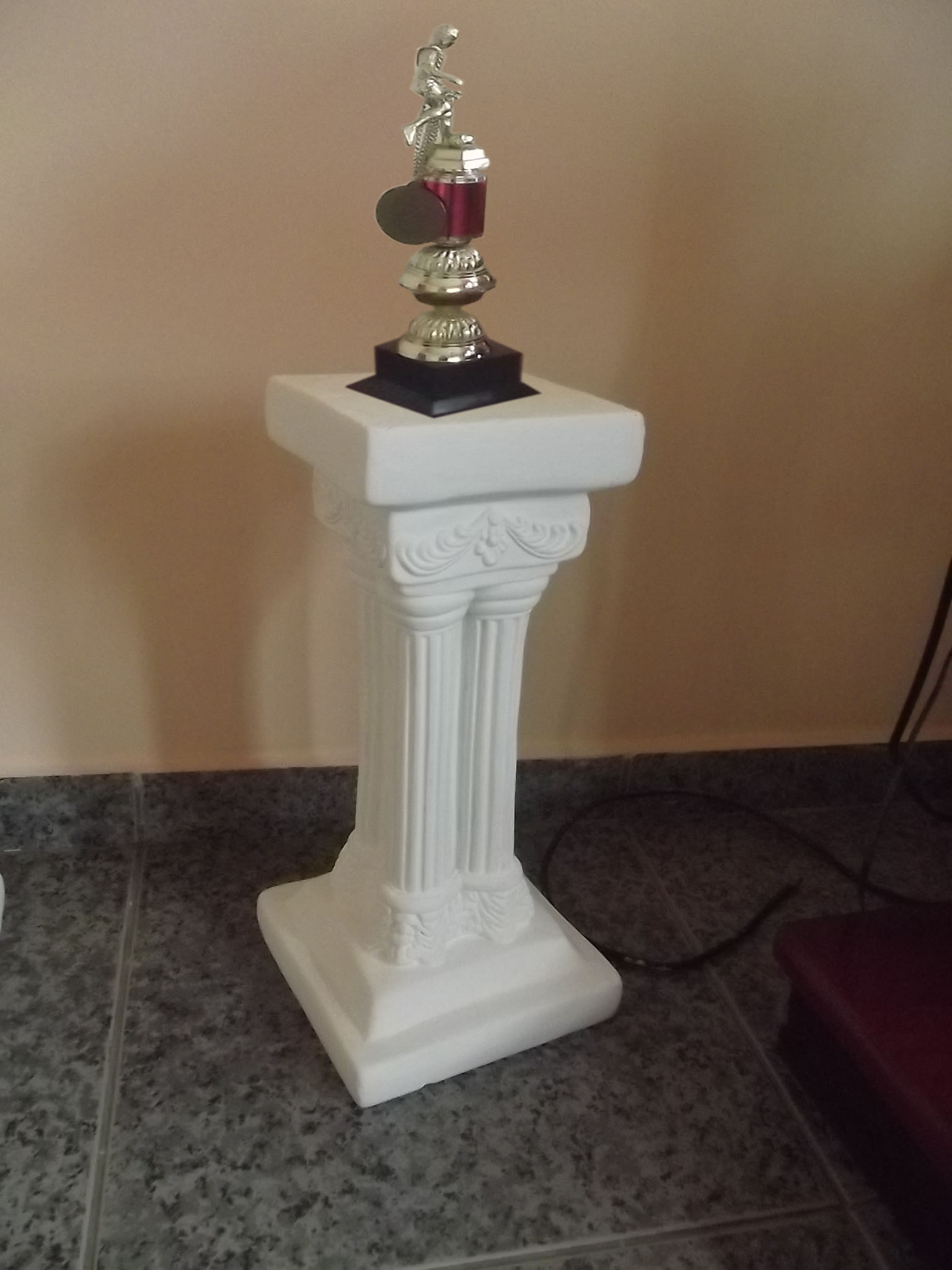}
\label{fig:HintExample1}}
\hfil
\subfloat[Ground truth edition mask.]{\includegraphics[width=1.5in]{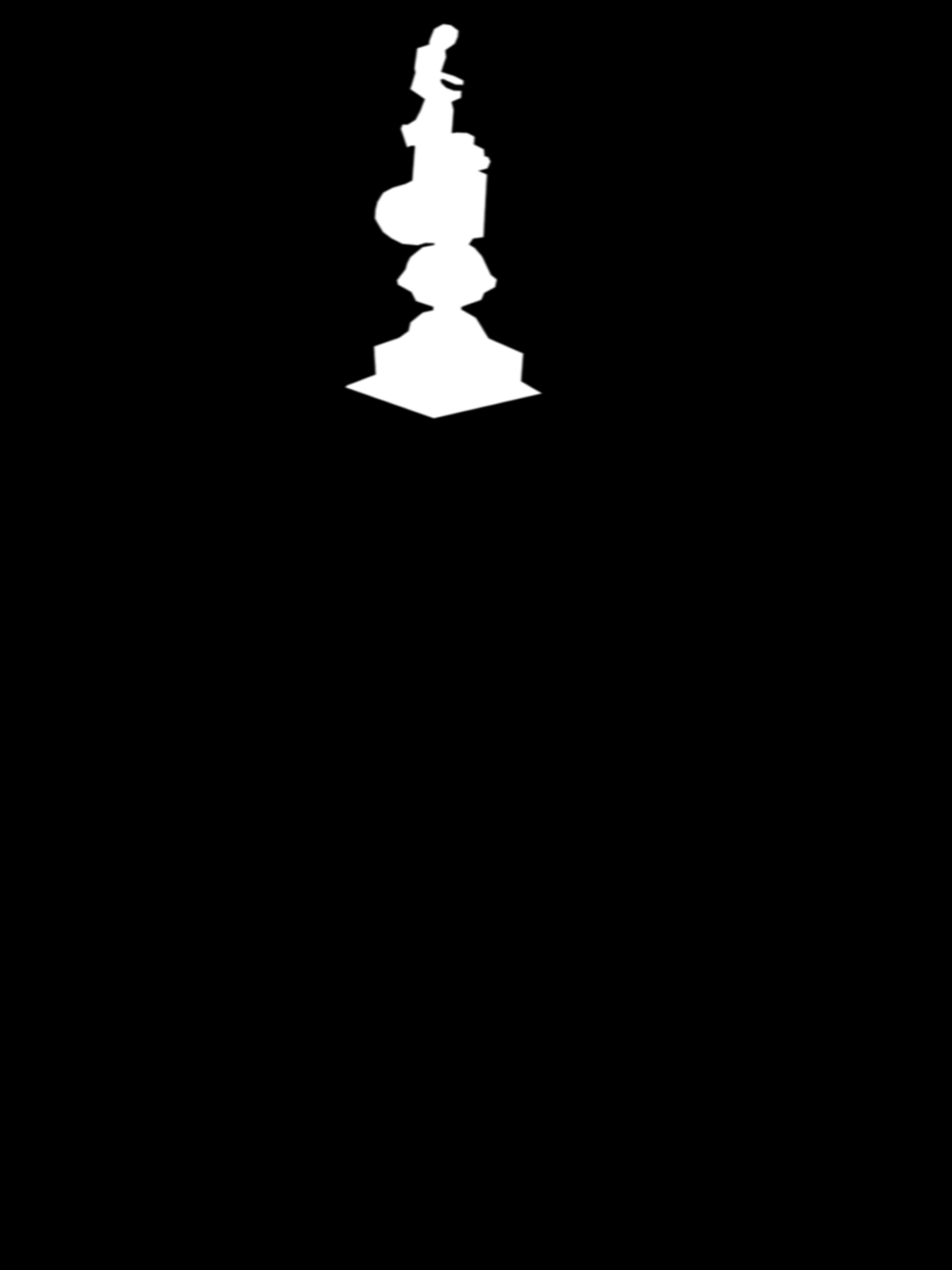}
\label{fig:HintExample2}}
\hfil
\subfloat[Evidence evaluation mask.]{\includegraphics[width=1.5in]{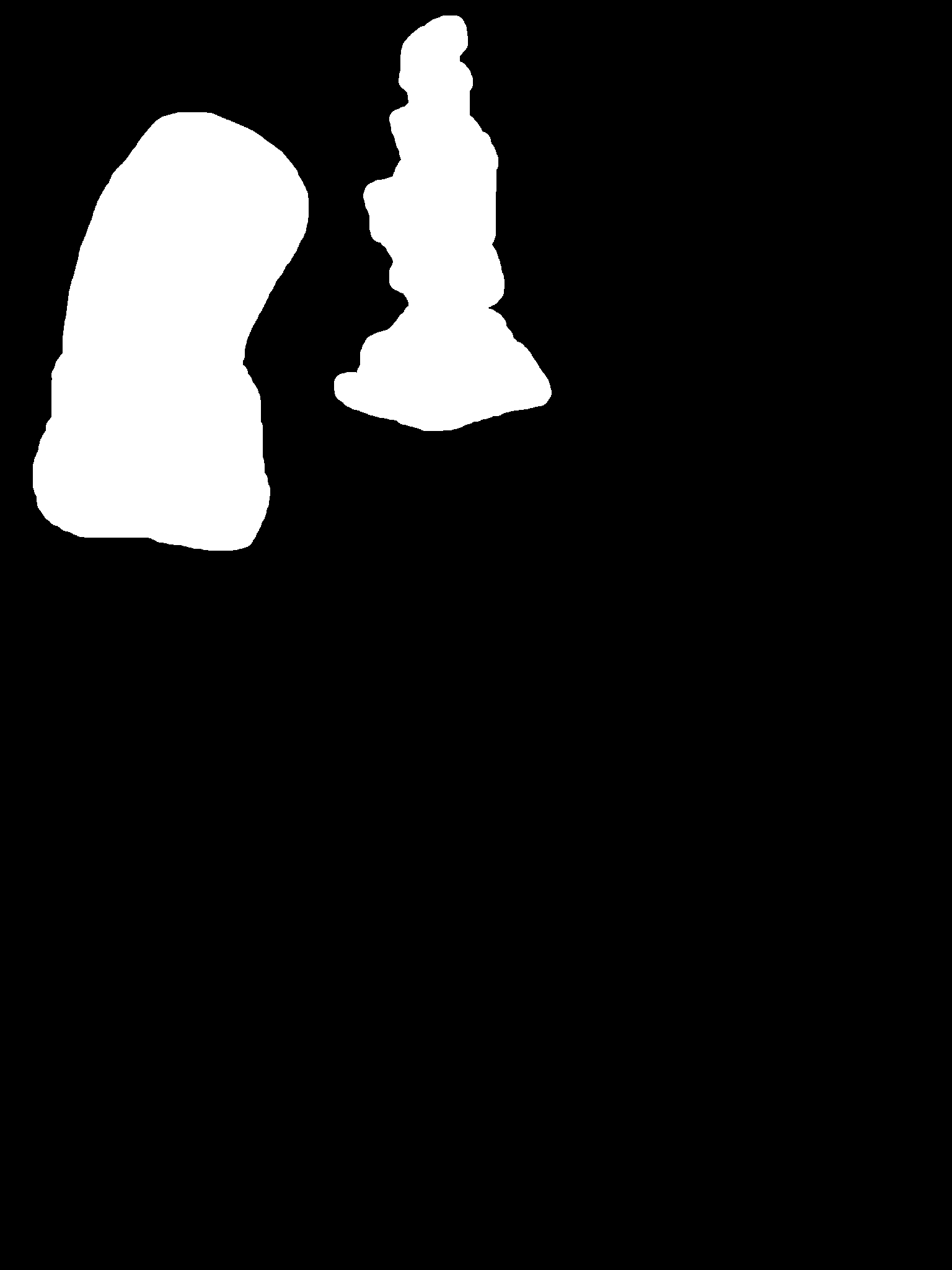}
\label{fig:HintExample3}}
\hfil
\subfloat[Image with hint.]{\includegraphics[width=1.5in]{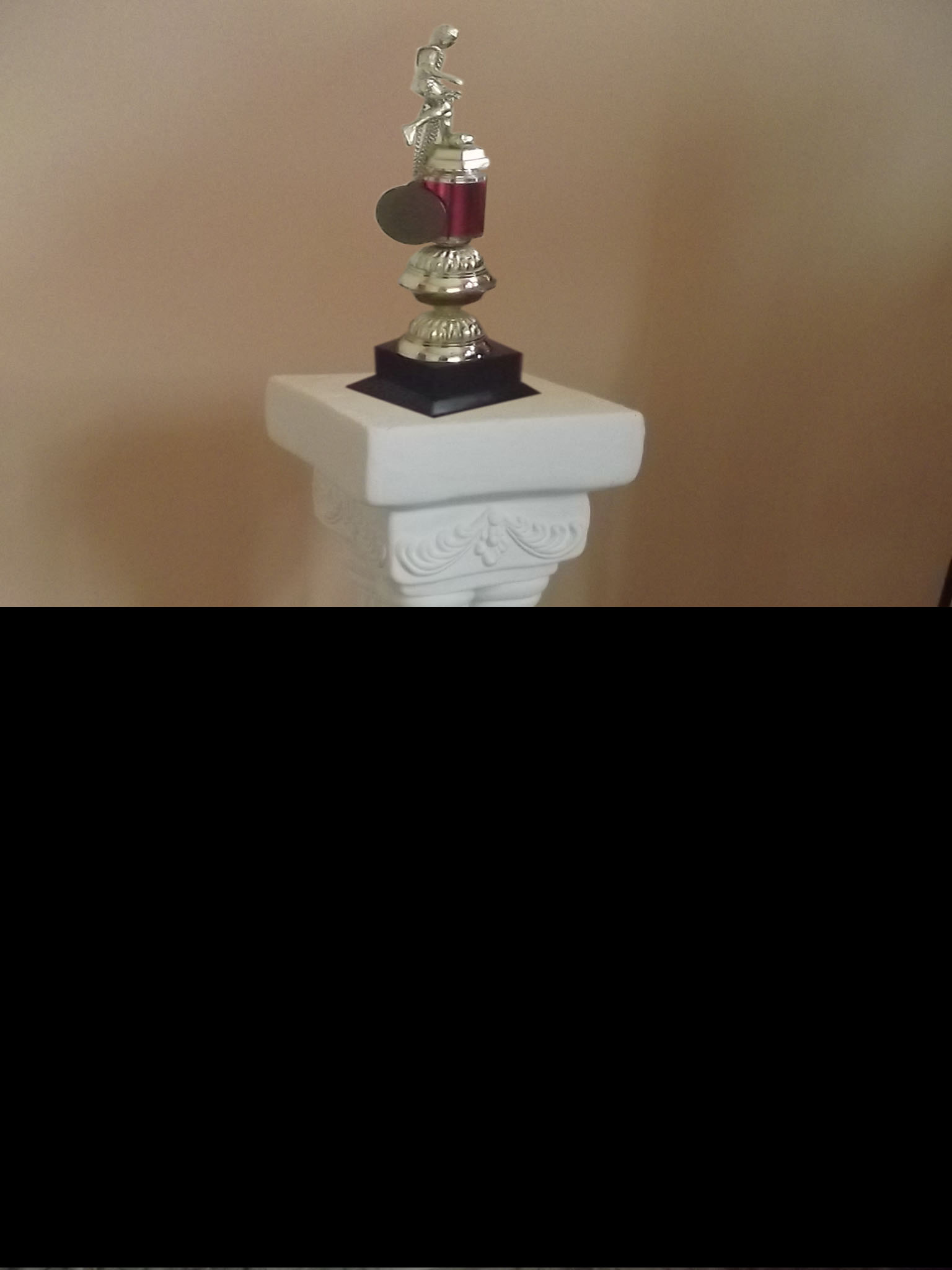}
\label{fig:HintExample4}}
\caption{Demonstration of the hint provided to the user for an image and the masks used in the process. In this case, the trophy was added to the pedestal. From left to right, the second picture is the ground truth edition mask, depicting the exact pixels that were changed in the manipulation. The third picture is a manual mask drawn over the ground truth mask that delimits all the regions in the image that can be considered valid evidence. In this case, the trophy, its surroundings and the shadow area. The figure on the right depicts the activated hint, blackening out a part of the image irrelevant to the process of finding manipulation.}
\label{fig:HintExample}
\end{figure*}

With each F image from the test database, we have associated a binary mask pointing to the forgery location and valid evidence, which is called the \textbf{evidence evaluation mask}. This mask is used for two purposes: to evaluate if the evidence provided by the user is valid or not; and to determine what part of the image can be discarded to provide a hint to the user. The evidence evaluation masks have been created using, as source, the ground truth binary masks of pixels changed in the doctoring process of each image. To cover different kinds of evidence users could provide, the masks were edited by hand increasing the valid area.

As can be seen in Figure~\ref{fig:HintExample}, the ground truth mask on Fig.~\ref{fig:HintExample2} only depicts the trophy added to the image. An observant user, however, might see the lacking trophy's shadow over the pedestal and use it as evidence to argue that the image is false. Another user might find the edges of the trophy to be irregular and click on a point in the outline of the trophy, but only slightly off the mask. Both answers, pointing shadow or edge problems, should be considered valid for justifying why the image is false, which means the ground truth mask alone is not sufficient for evaluating user provided evidence.

Figure.~\ref{fig:HintExample3} is the evidence evaluation mask for the manipulated image on Figure.~\ref{fig:HintExample1}, considering the points aforementioned. Since the masks are conditioned to subjective aspects of evaluation it is not a trivial task to achieve an ideal mask. In the case depicted in Figure.~\ref{fig:HintExample}, for instance, it could be argued that all of the pillar shadow should be considered valid evidence. This would increase the area of valid evidence, and consequently the chance of an user to guess right by chance, which is an important aspect to be taken in consideration for balancing purposes. No objective criteria was used for this balancing, as each mask is very dependent on the images' context.

The evidence evaluation masks are also used to generate the hint images, which allow the user to focus on the most relevant half of the image (Figure~\ref{fig:HintExample4}). The user can use one hint for image. In the case of a true image, a random half portion of the image is discarded. Hints are an important part of our user study because they can be used to determine two things: the impact of reducing the area of focus on the user performance, and which images can be considered harder, or more troublesome.

Considering all features, an user answer in our test is composed by seven elements: (1) user answer, either \textit{Yes, this image was manipulated} (F) or \textit{No, this image was not manipulated} (T); (2) answer confidence, either \textit{low}, \textit{medium} or \textit{high}; (3) user observation time before asking for a hint; (4) user observation time after asking for a hint; (5) if the user request a hint or not; (6) if the user viewed the image in its original resolution; (7) additional aspects of the image that the user found suspect: \textit{illumination}, \textit{shadows}, \textit{perspective}, \textit{geometry}, \textit{borders}, \textit{colors} or \textit{context}.

\subsection{The Database}

Our image database consists of 177 images, divided into 80 (45\%) true images and 97 (55\%) false images. The false images are split between 20 erasing images, 35 copy-and-paste images, and 42 splicing images. They have been handpicked from three public forensics image databases: the forensics challenge database\footnote{http://ifc.recod.ic.unicamp.br/fc.website/index.py}, the splicing database provided by \etal{Carvalho}~\cite{Tiago2013}, and \etal{Cozzolino}~\cite{Cozzolino2014} copy-and-paste database. The total image count adding all databases is around 6,000 images, with a great majority being true images, or false images with splicing operations.

\begin{figure}[ht]
\centering
\subfloat[T image.]{\includegraphics[width=0.22\textwidth]{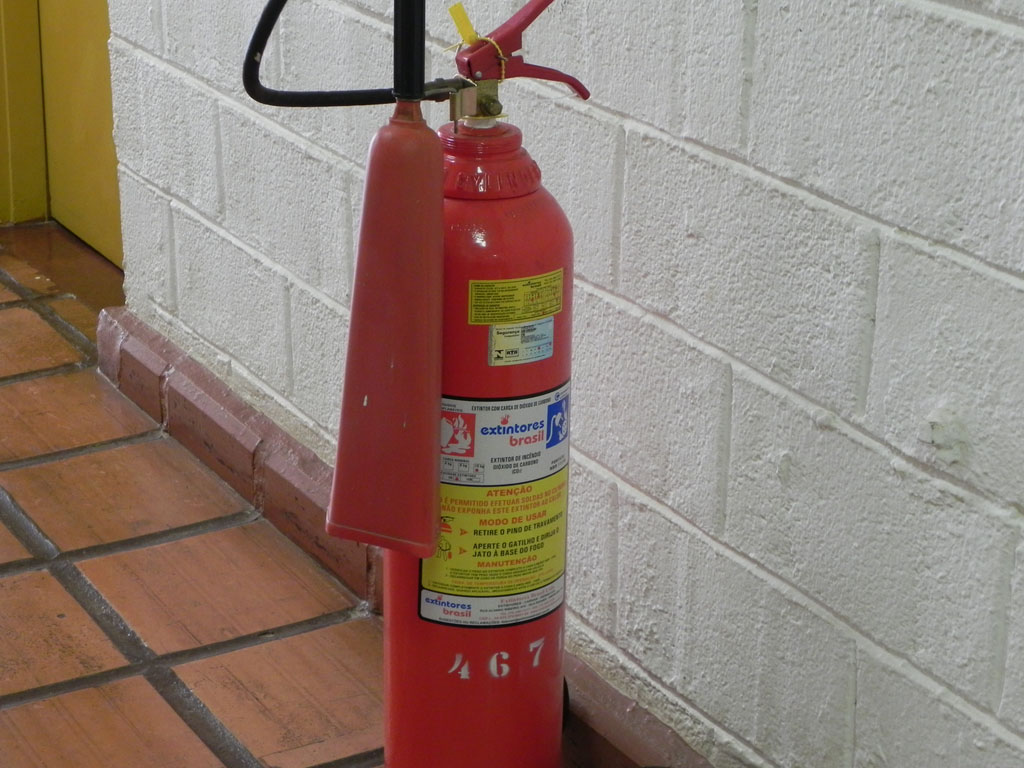}
\label{fig:true}}
\hfil
\subfloat[F image with an erasing forgery. Papers have been erased from the board.]{\includegraphics[width=0.22\textwidth]{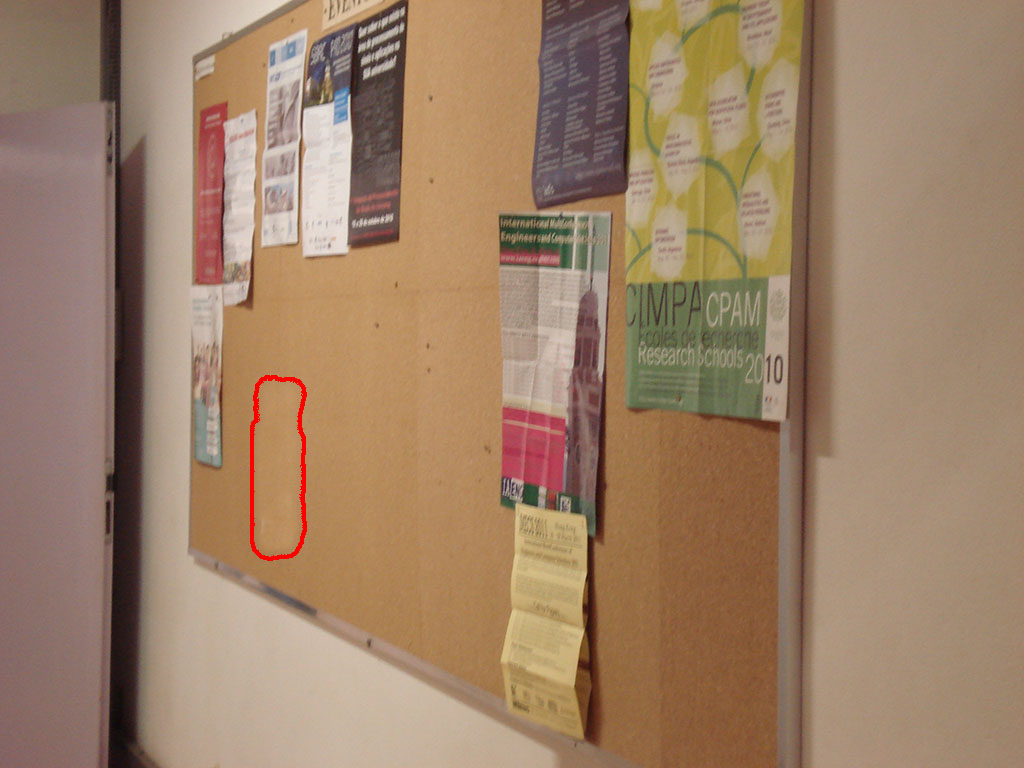}
\label{fig:erasing}}
\\
\subfloat[F image with a copy-paste forgery. One of the cisterns has been duplicated.]{\includegraphics[width=0.22\textwidth]{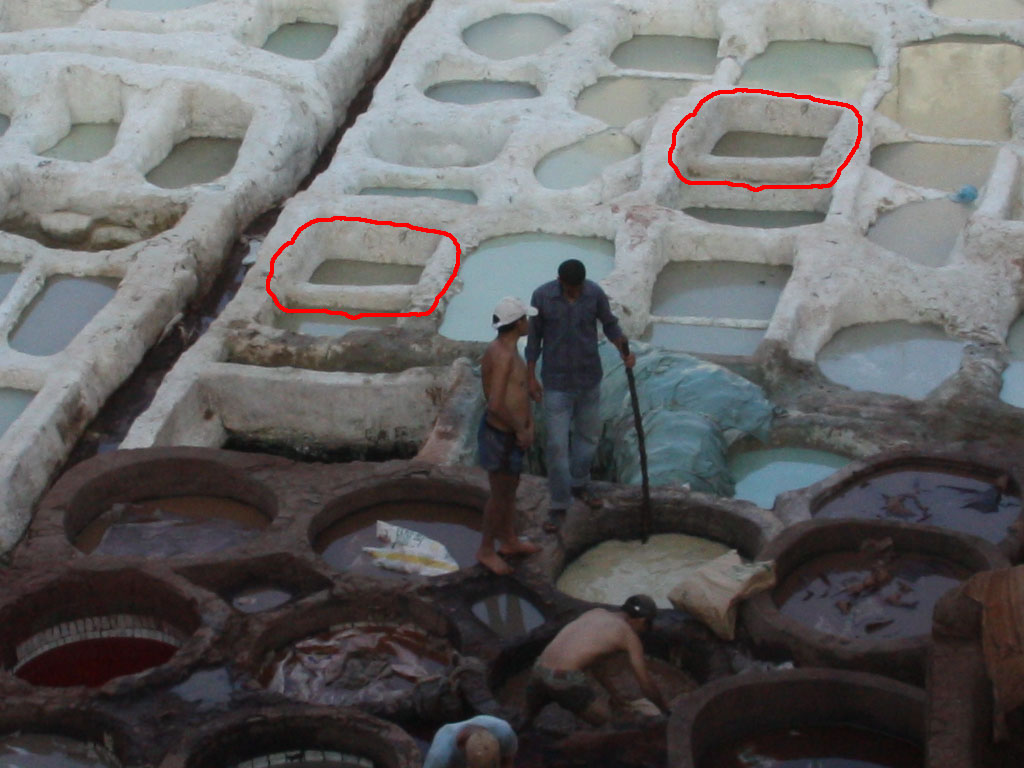}
\label{fig:copypasting}}
\hfil
\subfloat[F image with a splicing forgery. The man's face has been spliced from a different image.]{\includegraphics[width=0.22\textwidth]{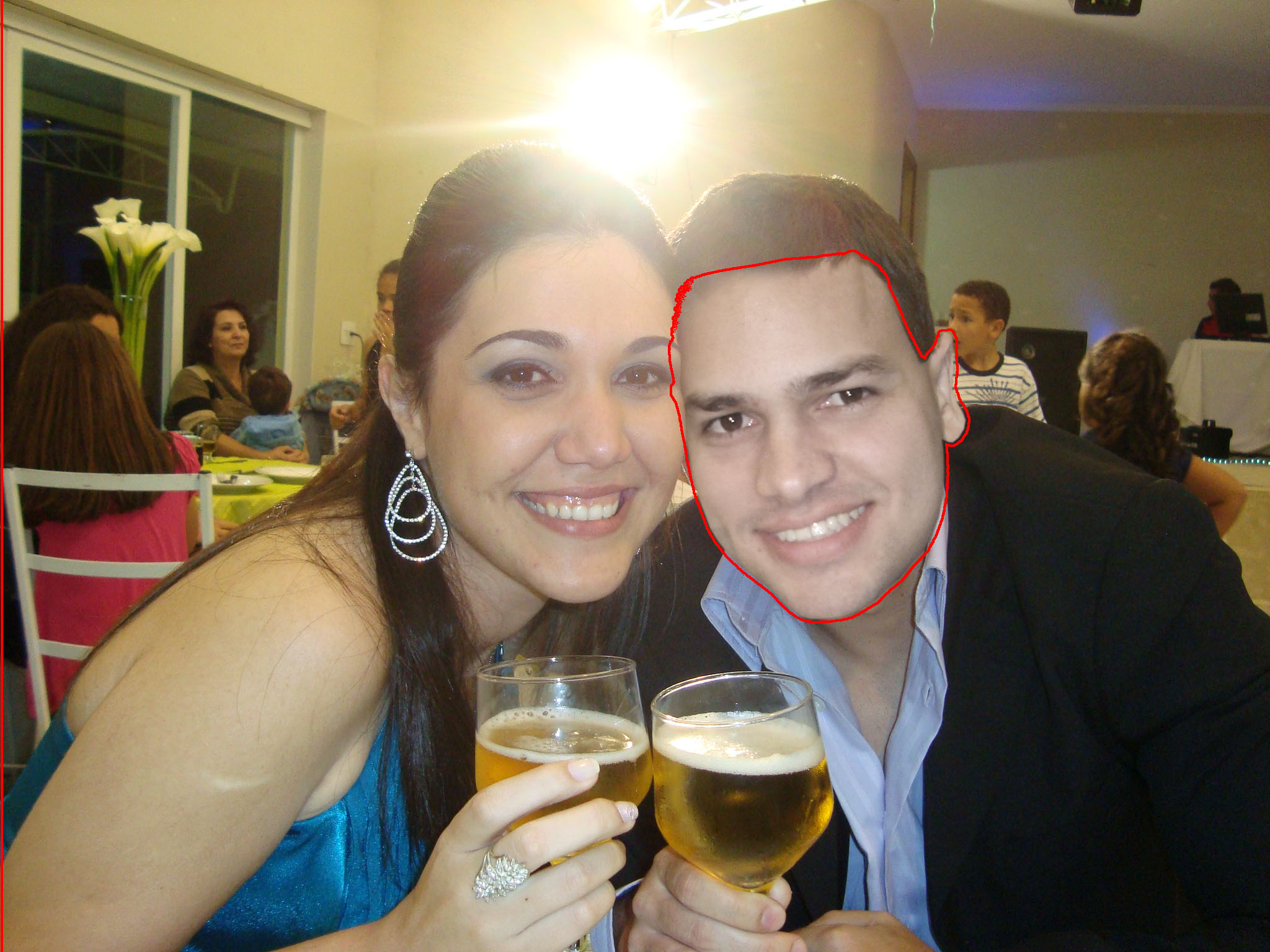}
\label{fig:splicing}}
\caption{Examples of different image types present in the assembled database. The edited area of F images is outlined in red. Here, we considered as erasing forgeries images where some region has been hidden by using brushes, blurring or even copying some small patches to cover it. A Copy-Paste forgery is when a region or object in an image is copied and pasted on the same image, with or without transformations such as scaling and rotation. Finally, a splicing forgery consists of copying a region from an image and pasting over another image, also with the possibility of transformations.}
\label{fig:databaseexamples}
\end{figure}

The motivation behind using known and public forensics databases was to avoid the bias of performing a test with self-made forgeries, and  to use images that are commonly analyzed by forensics techniques for comparison. The initial goal was to obtain a sample size of images between 150 and 200. To reach the final number of 177, an iterative process was done selecting images from the original pool into subsequently smaller pools. The following criteria were used:

\begin{itemize}
\item \textbf{Image type}: if it was a true image, erasing, copy-and-paste, or splicing forgery. This allows us to compare the results between different image types. Figure ~\ref{fig:databaseexamples} depicts examples of each type of evaluated forgery;
\item \textbf{Image context:} if the image depicted nature, people, buildings, landscapes, and if it was taken indoors or outdoors. Using this criteria, we achieved two main points: generalize our results in a way to cover a wide set of test  scenarios, and keep users interested during tests execution;
\item \textbf{Expected image difficulty:} how hard it would be for a subject to analyze the image. This was evaluated by the authors both subjectively and objectively, by personally inspecting forgeries;
\item \textbf{Edited area:} most images with multiple or large edited regions were considered inappropriate for the test, because they would conflict with the designed hint feature.
\end{itemize}

The manual database creation process is important because these three databases were designed mostly for non-assisted forensics techniques, resulting in a large number of forgeries that are not meaningful for users. An example of this would be copying and pasting two regions with the same color one over another. It might be trivial for a forensic technique that evaluates PRNU\footnote{Photo Response Non-Uniformity, a form of noise pattern for photo sensors.}~\cite{Chierchia2014}~\cite{Chierchia2013} or compression artifacts~\cite{Bianchi2012}~\cite{Qu2014} to identify this type of forgery, but it makes little sense to ask for a human user to do so. The small erasing images (just 20 images comprising this kind of forgery) is an unfortunate result of the lack of this type of image on the available public databases.

\subsection{User Motivation and Usability}

It is not easy for a user to analyze and answer all the 177 images in our study. The task can quickly become boring and underwhelming after a few dozen images. This is a serious problem that guided the design process of the study, and two main approaches were used to tackle it: providing motivation for users to go on, and ensuring each image answer could be treated equally, independently of the answering user and how many images he analyzed in total.

To motivate users to finish the study, serious games elements~\cite{ritterfeld2009} were used. The answering process was divided into 10 different levels, and at each provided answer the user gathered experience points to progress to a next level. The overall progress was displayed on the interface\footnote{Bar on the top right menu at Figure~\ref{fig:SiteExample1}} for users to easily keep track. Upon finishing a level, the user performance was calculated for the answered images on that level, with statistics such as hit rate, average time, confidence, and hint usage. 

These statistics are based on all the images answered on the level finished. It is not possible for the user to determine exactly which images were answered right or wrong. This information is saved as part of the users' profile and can be reviewed at any time, summarizing their performance for each finished level. This is an important feature, because users need constant feedback, and providing information on each answer they got right or wrong would compromise the study methodology.

Upon finishing all levels and completing the test, a user earns the right to appear on the high scores page\footnote{http://newton.inf.ufrgs.br/scores.php}, where his overall statistics are displayed. This feature was added as a direct suggestion from early testing users\footnote{These users were not included on the final answer pool.} that argued they felt motivated by the competition aspect. It is important to note that the level progression system does not affect, in any way, the order of images, nor it presents any change in difficulty. It serves merely as a motivational and progress keeping mechanism.

The order in which the images appear for each user is semi-random, based on an algorithm that prioritizes the images that have received less answers to be picked more often. When a user loads the main test page, this algorithm randomly selects an image from the pool of least answered ones, and reloading the page or coming back another time evokes this process again, changing the image. Once an image has been answered by the user, it cannot be selected again for him. The reason for this is to guarantee that all images have a similar amount of answers, so the collected data is as homogenous as possible regarding to answer distribution.

The user interface was designed to be as user friendly and consistent as possible. It was tested on the most common browser resolutions\footnote{$1336\times768$, $1280\times800$ and $1920\times1080$}, displaying the image in fixed resolution of $1024\times768$ for all users. The fixed resolution is important for consistency, but all images can be opened in their original resolutions by clicking in the link under it. Which images the user opened in full resolution is also one of the features tracked in our study.

Since the study is based on visual aspects and can be exhaustive for users, special effort was made to minimize stress and confusion. Several iterative development cycles were done on the interface, and features that were tweaked included but were not limited to: colors, button size and placement, text, menu size and placement, and tooltips. All textual information provided is bilingual, both in English and Brazilian Portuguese.

Further discussion and validation of our approach is provided in  Appendix~\ref{sec:influenceLevel}.

%% file: sec/results.tex
\section{Results and Discussion}
\label{sec:Exposition}
In this section we analyze the results obtained from the collected data. The users' answers are classified according to the criteria shown in Table.~\ref{tab:answerClasses}. 

To determine if two features are correlated, the Pearson coefficient is
estimated for both $\rho >0$ (positively correlated) and $\rho <0$
(negatively correlated). We estimated the $p-$value, which can be defined as
the \textbf{smallest choice} that could be done for the significance level ($%
\alpha $) of the statistical test (i.e., the false positive error, or simply
the probability of a incorrect rejection of a true null hypothesis), for that
we reject the null hypothesis. In other words, if the $p$-value is smaller than the significance level, we must reject the null hypothesis, and the correlation between the features is significant~\cite{trivedi2002probability}. 

The next subsection presents an overview of all collected data, followed by a more in-depth analysis of users and images statistics.

\subsection{Overview}
\label{sec:Overview}

\begin{figure}[ht]
  \centering
  \includegraphics[width=3in]{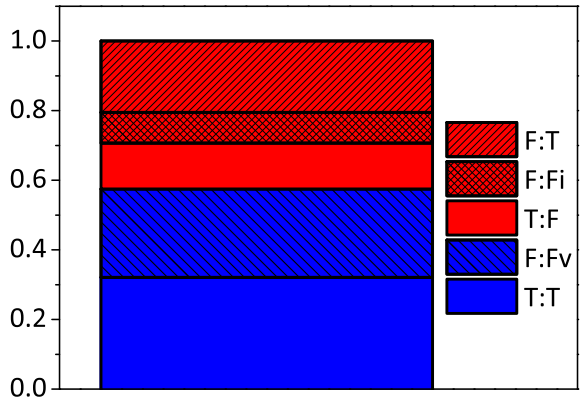}
  \caption{Overview of the class distribution among all answers.}
	\label{fig:percentOverview}
\end{figure}

Figure~\ref{fig:percentOverview} shows the overall distribution of classes between all answers. We collected 17,208 answers from 393 different users, after discarding invalid entries. The dark and light blue bars represent the correct answer classes (T:T and F:Fv, respectively). In this visualization, one can see the accuracy (sum of the bottom blue classes), and the proportion of a particular class.

A more detailed account of the answers can be seen on Table~\ref{tab:answerOverview}. In the center column, the amount of individual answers for that particular category is displayed, and on the right the proportion relative to the answer pool. On this table there is a distinction between images that were ``answered correctly" or ``classified correctly". 

A correctly classified image is when a user guessed if it was a T or an F image, regardless of the evidence provided (classes T:T, F:Fv, F:Fv, and F:Fi). Images answered correctly are the ones where the user guessed the type right, \textbf{and provided valid evidence for the false ones} (classes T:T, and F:Fv only). This distinction is important because the evidence evaluation was performed offline, after the answers were collected. All feedback the users received on their hit rates (upon completing a level, for instance) and the information presented on the website is regarding images that were correctly classified, but not necessarily correctly answered.

\begin{table}
		\centering
    \begin{tabular}{|l|c|c|}
    \hline
    ~                          & Total & Proportion \\ \hline
    T Images Answered       & 7,791  & 0.452      \\
    F Images Answered      & 9,471  & 0.548      \\ \hline   
    T Answers               & 9,048  & 0.525      \\
    F Answers              & 8,160  & 0.475      \\ \hline
    Classified Correctly           & 11,409 & 0.663      \\
    Classified Wrongly           & 5,799  & 0.337      \\ \hline
    Answered Correctly             & 9,899  & 0.576      \\
    Answered Wrong             & 7,309  & 0.424      \\ \hline
		Erasing Images Answered    & 1,942  & 0.113      \\
    Copy-Paste Images Answered & 3,392  & 0.197      \\
    Splicing Images Answered   & 4,083  & 0.237      \\ \hline
		T:T \ \ Answers            & 5,520  & 0.320      \\
		F:Fv \ Answers             & 4,379  & 0.254      \\
		T:F \ \ Answers            & 2,271  & 0.132      \\
		F:Fi Answers               & 1,510  & 0.087      \\
		F:T \ \ Answers            & 3,528  & 0.205      \\
		\hline
    \end{tabular}
		\caption{Distribution of answers. From top to bottom, each row group divides the total 17,208 answers according to different criteria. The first group accounts for the type of images being answered. The second group represents the types of answers given by the users, without considering the image type or evidence. The third and fourth groups relate to which images were correctly classified or answered, respectively. The fifth group separates answers according to the type of forgery on the image, and the final group is a relation of all the answer classes.}
		\label{tab:answerOverview}
\end{table}

\begin{table}[]
\centering
\begin{tabular}{cc|c|l|c|l|c}
\cline{3-6}
                                              &                    & \multicolumn{4}{c|}{Image}                                                   &                                             \\ \cline{3-6}
                                              &                    & \multicolumn{2}{c|}{F}                 & \multicolumn{2}{c|}{T}              &                                             \\ \hline
\multicolumn{1}{|c|}{\multirow{2}{*}{Answer}} & \multirow{2}{*}{F} & \multicolumn{2}{c|}{4,379 (F:Fv)}       & \multicolumn{2}{c|}{2,271 (T:F)}     & \multicolumn{1}{c|}{\multirow{2}{*}{6,650}}  \\ \cline{3-6}
\multicolumn{1}{|c|}{}                        &                    & \multicolumn{2}{c|}{True Positive}     & \multicolumn{2}{c|}{False Positive} & \multicolumn{1}{c|}{}                       \\ \cline{2-7} 
\multicolumn{1}{|c|}{}                        & \multirow{2}{*}{T} & \multicolumn{2}{c|}{5,038 (F:T + F:Fi)} & \multicolumn{2}{c|}{5,520 (T:T)}     & \multicolumn{1}{c|}{\multirow{2}{*}{10,558}} \\ \cline{3-6}
\multicolumn{1}{|c|}{}                        &                    & \multicolumn{2}{c|}{False Negative}    & \multicolumn{2}{c|}{True Negative}  & \multicolumn{1}{c|}{}                       \\ \hline
                                              &                    & \multicolumn{2}{c|}{9,417}              & \multicolumn{2}{c|}{7,791}           &                                             \\ \cline{3-6}
\end{tabular}
\caption{Confusion matrix for user answers, considering a binary classification problem. The horizontal axis corresponds to the image type (T or F), and the vertical axis to the answer given by the user (also T or F). Since the context of the classification problem is to identify the false images, a true positive corresponds to correctly identifying a false image, and so on. F:Fi is grouped together with F:T because we consider it to be a false negative. The image is F, but the user provided wrong evidence, so he was not able to identify the actual altered region.}
\label{tab:confusionMatrix}
\end{table}

Using the data provided on Table~\ref{tab:answerOverview} and Table~\ref{tab:answerClasses}, we construct a confusion matrix to further evaluate the binary classification. Note that the false negative class corresponds not only to failing to identify an F image (F:T), but also to providing invalid evidence (F:Fi). The confusion matrix is presented on Table~\ref{tab:confusionMatrix}, and the classification statistics are displayed on Table~\ref{tab:classificationStats}.

\begin{table}[h]
\centering
{\renewcommand{\arraystretch}{1.7}
\begin{tabular}{|l|c|c|}
\hline
Statistic    &  Formula & Value \\ \hline
Accuracy     &  $\frac{F:Fv+T:T}{F:Fv+F:Fi+T:T+T:F+F:T}$ & 0.575 \\ \hline
Precision    &  $\frac{F:Fv}{F:Fv+T:F}$ & 0.658 \\ \hline
Specificity  &  $\frac{T:T}{T}$  & 0.708 \\ \hline
Sensitivity  &  $\frac{F:Fv}{F}$ & 0.465 \\ \hline
F1 Score     &  $\frac{2(F:Fv)}{2(F:Fv)+T:F+F:Fi+F:T}$ & 0.545 \\ \hline
\end{tabular}
}
\caption{Classification statistics for the users answers using the confusion matrix on Table~\ref{tab:confusionMatrix}. Accuracy is the amount of correct answers over total answers. The precision is the amount of true positive answers over all true positive and false positive answers. Specificity is the amount of true negative answers over all negative answers. Sensivity is the amount of true positive answers over all positive answers. The F1 score is the harmonic mean of precision and sensivity.}
\label{tab:classificationStats}
\end{table}

The most important observation is that people are not good at identifying which images were altered and which were not, with an accuracy of 0.575. A sensitivity value of 0.465 means that of all the F images answered, only 46.5\% were guessed right. A higher value of specificity than sensivity indicates that people tend to answer images are T much more often. This is also corroborated by the data on Table~\ref{tab:answerOverview}: while the majority of images that appear on the test were F rather than T (9,471 against 7,791), users provided more T answers (9,048 against 8,160). This is understandable, as providing a T answer requires no additional evidence. When in doubt, users defaulted to a T answer. 

The values in Table~\ref{tab:classificationStats} are very low if compared to forensic techniques. Considering only works that have images in common with our database, \etal{Cozzolino}~\cite{Cozzolino2014} and \etal{Chierchia}~\cite{Chierchia2014} have consistently achieved an F1 Score over 0.8 for the copy-paste images; \etal{Carvalho}~\cite{Tiago2013} best configuration for spliced images reaches up to 0.68 sensitivity and 0.9 specificity. The Ranking of the International Forensics Challenge~\footnote{http://ifc.recod.ic.unicamp.br/fc.website/index.py?sec=11} uses a different metric, but shows that the best contestants achieved a nearly perfect identification rate.

\subsection{User Background and Behavior}
\label{sec:UserBackground}

\begin{figure}[t]
  \centering
  \includegraphics[width=3in]{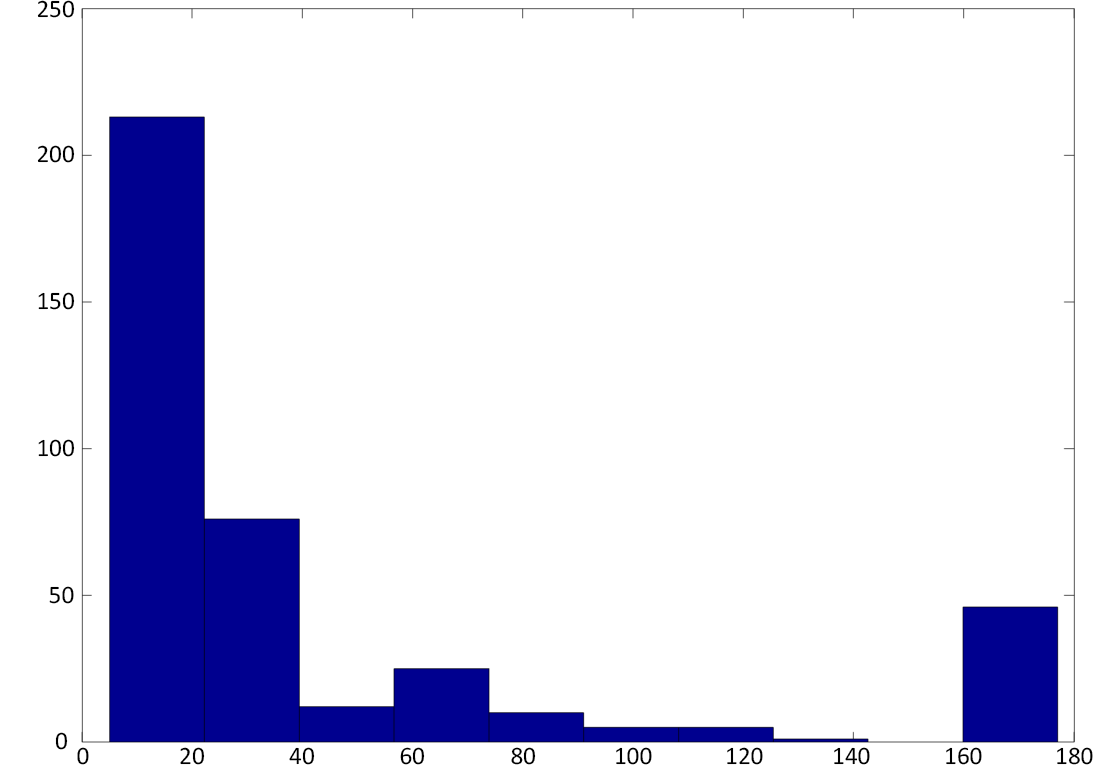}
  \caption{Histogram of user distribution per amount of answers. The X axis represents the amount of images a user has answered, and the Y axis the amount of users that have answered that amount.}
	\label{fig:userAnswerHist}
\end{figure}

Our user study was designed to maximize the amount of total individual answers. For this purpose, all answers are treated equally, independent of user. The histogram in Figure~\ref{fig:userAnswerHist} shows the number of questions answered by the participants. The majority of users did not answer more than 40 images, but over 50\% of the answers gathered were provided by users who answered more than 100 images. In total, 46 users fully completed the test by answering all 177 images.

\begin{table}
		\centering
    \begin{tabular}{|l|c|c|}
    \hline
    ~                          & Total & Proportion \\ \hline
    Age                        &       &            \\ \hline 
    Up to 21                   & 513   & 0.029      \\   
    21 to 35                   & 14,738 & 0.856      \\
    35 to 50                   & 1,359  & 0.079      \\ 
    50 and Over                & 598   & 0.034      \\ \hline
		Education                  &       &				   	\\ \hline
		Highschool                 & 791   & 0.046      \\
		Undergraduate              & 7,123  & 0.413      \\
		Graduate                   & 9,294  & 0.540      \\ \hline
		Experience                 &       &				   	\\ \hline
		User                       & 4,621  & 0.268      \\
		Amateur                    & 9,537  & 0.554      \\
		Professional               & 3,050  & 0.177      \\ \hline
    \end{tabular}
		\caption{Overview of the distribution of answers according to user backgrounds.}
		\label{tab:answerBackground}
\end{table}

We treat user information (i.e. age, education, and experience) as answer features to quantify the contribution of each background to the answer pool. This information is displayed in Table~\ref{tab:answerBackground}. The majority of answers were provided by adults from 21 to 35 years of age, with graduate education and an amateur level of experience with digital images. This particular background configuration arguably represents one of the best suited demographics to perceive manipulation in digital images, so it is possible that it represents the upper bounds of the general population in terms of accuracy.

To better understand how user features and accuracy are related, we estimated the Pearson correlation. Age, Education and Experience are considered background, and the following features determine user behavior:

\begin{itemize}
\item Confidence: the average confidence between all the users' answers;
\item Time: the average time elapsed before the user asked for a hint in all answers;
\item Time after Hint: after asking for a hint, how much time on average the user analyzed the image before answering;
\item Hint Proportion: the proportion of answered images in which the user asked for a hint;
\item Full Resolution: the proportion of answered images in which the user opened the image in full resolution.
\end{itemize}

The correlation between the user accuracy, each answer class in particular (T:T, F:Fv, F:T, F:Fi, T:F) and the above features are displayed in Table~\ref{tab:corrUserFeaturesClasses}. The first column for each pairing is the Pearson's $\rho$ value, and the second is the $p$. Significant correlations ($p<0.05$) are highlighted in blue when $\rho>0$, meaning a positive correlation, and in red when $\rho<0$, for negative correlations.

\begin{table*}[t]
\centering
    \begin{tabular}{|l|ll|ll|ll|ll|ll|ll|}
    \hline
    ~                & Acc. $\rho$ & Acc. $p$ & T:T $\rho$    & T:T $p$    & F:Fv $\rho$  & F:Fv $p$   & T:F $\rho$   & T:F $p$    & F:Fi $\rho$  & F:Fi $p$  & F:T $\rho$   & F:T $p$    \\ \hline
    Confidence  &\ \color{Blue}{0.143}&\color{Blue}{0.004}&\ 0.085& 0.091&\ 0.080&0.109&\color{BrickRed}{-0.159}&\color{BrickRed}{0.001}&-0.0315&0.533&-0.011& 0.819    \\
    Time        & \ 0.033      & 0.503       & \color{BrickRed}{-0.188}&\color{BrickRed}{1e-04}& \ \color{Blue}{0.228}&\color{Blue}{5e-06}& \ \color{Blue}{0.197}&\color{Blue}{8e-05}& \ \color{Blue}{0.104}&\color{Blue}{0.039}&\color{BrickRed}{-0.288}&\color{BrickRed}{6e-09}\\
    Time after Hint   & -0.051       & 0.306       &\color{BrickRed}{-0.136}&\color{BrickRed}{0.006}& \ 0.076    & 0.131    & \ \color{Blue}{0.190}&\color{Blue}{1e-04}&\ \color{Blue}{0.135}&\color{Blue}{0.007}&\color{BrickRed}{-0.200}&\color{BrickRed}{8e-05} \\
    Hint Proportion    & -0.019       & 0.704       &\color{BrickRed}{-0.125}&\color{BrickRed}{0.013}& \ \color{Blue}{0.103}&\color{Blue}{0.040}&\ \color{Blue}{0.150}&\color{Blue}{0.002}& \ \color{Blue}{0.102}&\color{Blue}{0.042}&\color{BrickRed}{-0.018}&\color{BrickRed}{3e-04}\\
    Full Resolution       & \ \color{Blue}{0.116}&\color{Blue}{0.021}& -0.005     & 0.919    & \ \color{Blue}{0.139}&\color{Blue}{0.005}& \ 7e-04   & 0.988    &\color{BrickRed}{-0.122}&\color{BrickRed}{0.015}& -0.058    & 0.248    \\
    Age         &\color{BrickRed}{-0.146}&\color{BrickRed}{0.003}& -0.082     & 0.102    & -0.086     & 0.087    & \ 0.064   & 0.200    & \ 0.079   & 0.115   & \ 0.066   & 0.185    \\
    Experience  & \ \color{Blue}{0.122}&\color{Blue}{0.015}& \ 0.051    & 0.310    & \ 0.089    & 0.076    & -0.040    & 0.427    & -0.023    & 0.638   & -0.097    & 0.054   \\
    Education   &-0.035       & 0.479       & -0.034     & 0.493    & -0.006     & 0.895    & -0.010    & 0.841    & \ 0.033   & 0.507   & \ 0.030   & 0.552    \\ \hline
    \end{tabular}
		\caption{Correlation and respective $p$-value between performance, answer classes and user features. Here, blue denotes positive correlation with acceptable $p$-value ($p<0.05$), and red denotes negative correlation with acceptable $p$-value. Black values do not satisfy the threshold and the null hypothesis cannot be rejected.}
		\label{tab:corrUserFeaturesClasses}
\end{table*}

There are significant correlations between confidence, time, age, experience, and the accuracy (proportion of T:T and F:Fv over all answers). This means that users with higher accuracy tend to analyze the image more carefully by opening it in full resolution more often, and are more confident in their answers. In terms of background, the experience is the only factor that can increase accuracy, while education has no meaningful effect. 

Both time features are correlated with providing F answers in detriment of T answers, regardless of the image type. What can be inferred from this is that users that spend too much time analyzing an image are prone to suspecting it is F, even if there is no clear evidence. We call this behavior \textit{over-analyzing}. The same can be observed on the Hint use, with a slightly smaller $\rho$ value. If a user is already suspecting an image to be F, the hint only increases his suspicion. This bias could be explained by the nature of the test, where users were exhaustively scrutinizing the images for irregularities. 

In Table~\ref{tab:corrUserFeatures}, the correlations between all the behavior and background features themselves are presented. The name of each feature is shortened to its initials in the table to reduce space use. While there is no direct correlation between education and accuracy, the former does increase the average confidence, and experience increases with age. Since age correlates negatively with accuracy and positively with experience, this indicates there is a plateau of accuracy for age, which was found to be from 24 to 30 years. Furthermore, users who tend to ask more for hints have lower confidence, and those who take more time analyzing the images are more prone to ask for hints and spend even more time looking at them.


\begin{table*}[t]
\centering
\begin{tabular}{|l|c c|c c|c c|c c|c c|c c|c c|}
\hline
 & T $\rho$ & T $p$ & TH $\rho$ & TH $p$ & H $\rho$ & H $p$ & FR $\rho$ & FR $p$ & A $\rho$ & A $p$ & Ed $\rho$ & Ed $p$ & Ex $\rho$ & Ex $p$ \\
\hline
C  & -0.031 & 0.542 &\color{BrickRed}{-0.147}&\color{BrickRed}{0.004}&\color{BrickRed}{-0.228}&\color{BrickRed}{5e-05}& 0.019 & 0.711 &\ 0.064 & 0.206 &\ \color{Blue}{0.164}&\color{Blue}{0.001}& -0.051 & 0.310 \\
T  &      - &      -&\ \color{Blue}{0.224}&\color{Blue}{7e-06}&\ \color{Blue}{0.195}&\color{Blue}{1e-04}&\color{Blue}{0.121}&\color{Blue}{0.016}&\ 0.080 & 0.112 & -0.003 & 0.952 & -0.011 & 0.831 \\
TH &      - &      -&     -  &     - &\ \color{Blue}{0.728}&\color{Blue}{4e-65}& \color{Blue}{0.125}&\color{Blue}{0.013}& -0.041 & 0.414 & -0.035 & 0.486 & -0.011 & 0.822 \\
HP &      - &      -&     -  &     - &     -  &     - &\color{Blue}{0.166}&\color{Blue}{0.001}& -0.042 & 0.405 & -0.090 & 0.075 &\ 0.031 & 0.539 \\
FR &      - &      -&     -  &     - &     -  &     - &     - &     - &\ 0.034 & 0.503 &\ 0.086 & 0.089 &\ 0.085 & 0.091 \\
A  &      - &      -&     -  &     - &     -  &     - &     - &     - &      - &     - &\ 0.028 & 0.578 &\ \color{Blue}{0.385}&\color{Blue}{2e-15}\\
Ed &      - &      -&     -  &     - &     -  &     - &     - &     - &     -  &     - &\     - &     - &\ 0.046 & 0.363 \\
\hline
\end{tabular}
\caption{Correlation and respective $p$-value between the different user features. T stands for Time, TH for Time after Hint, C for Confidence, H for Hint Proportion, FR for Full Resolution, A for Age, and Ed for Education. Here, blue denotes positive correlation with acceptable $p$-value ($p<0.05$), and red denotes negative correlation with acceptable $p$-value. Black values do not satisfy the threshold and the null hypothesis cannot be rejected.}
\label{tab:corrUserFeatures}
\end{table*}

\subsection{Images}
\label{sec:Images}

Answer features can be grouped by image for a more in-depth analysis. We also distinguish between image behavior and image background features. Image background features are measures of an image's properties, while behavior features relate to the behavior of users answering it. We estimate two features for T images and three features for F images:

\begin{itemize}
\item Luminance: the average luminance value for all pixels in the image using the formula $0.2126R+0.7152G+0.0722B$~\cite{luminance1996};
\item Variance: for all pixels, we computed the standard deviation of their 11x11 neighbors. The ``Variance" is the average of this value for all pixels.
\item Edited Area: for F images, the proportion of the edited area in pixels over the total resolution.
\end{itemize}

The main goal in observing images' background features is to analyze correlations between their intrinsic properties and how users answer them. The local variance on a 11x11 window was the feature chosen to measure an image's visual cluttering, which we called the Variance. Due to the images' different base resolutions, capture devices, and camera settings, which affect the local variance, we tested different window sizes (3x3, 5x5, etc. up to 17x17). A window size of 11x11 was the smallest that yielded a significant result ($p<0.05$), and as such was chosen to represent the Variance. To further reduce the effect of the difference in the images' resolutions, they were re-sized to the resolution used in the interface (1024x768), before estimating the image features. 

An image's behavior features are calculated by averaging the features of all its received answers, as follows:

\begin{itemize}
\item Confidence: the average confidence of all answers this image received;
\item Time: the average time elapsed before all users answering this image asked for a hint;
\item Time after Hint: after asking for a hint, how much time on average the users analyzed the image before answering;
\item Hint Proportion: the proportion of answers this image received in which a hint was asked for;
\item Full Resolution: the proportion of answers this image received in which it was observed in full resolution.
\end{itemize}

To estimate the correlation between image features and answering classes, it is necessary to differentiate between T and F images, because they have mutually exclusive classes\footnote{T:T, and T:F for T images, and F:Fv, F:Fi, and F:T for F images}. Note that for T images, the proportion of T:T answers is the accuracy, and the same holds for F images and F:Fv answers. The data is split between Table~\ref{tab:corrTrueImgs} for T images and Table~\ref{tab:corrFakeImgs} for F images. Since there are only two classes for T images, the correlations between T:T and T:F are complementary: they have the same $p$-value, and $\rho$ values with opposite signs. 

\begin{table}[h]
\centering
\begin{tabular}{|l|c c|}
\hline
 &T:T $\rho$ & T:T $p$ \\
\hline
Confidence & -0.062 & 0.584\\
Time & -0.106 & 0.350\\
Time after Hint & -0.200 & 0.075\\
Hint Proportion & -0.167 & 0.138\\
Full Resolution & \color{BrickRed}{-0.261} & \color{BrickRed}{0.020}\\
Luminance & 0.105 & 0.352\\
Variance & \color{Blue}{0.230} & \color{Blue}{0.040}\\
\hline
\end{tabular}
\caption{Correlation and respective $p$-value of image features for true images and the T:T class. The class T:F is omitted for its complementarity. Here, blue denotes positive correlation with acceptable $p$-value ($p<0.05$), and red denotes negative correlation with acceptable $p$-value. Black values do not satisfy the threshold and the null hypothesis cannot be rejected.}
\label{tab:corrTrueImgs}
\end{table}

For T images, only two significant correlations were found. The more users have opened images in full resolution, the more they tend to answer that the respective images are F. This seems to contradict the notion that users who analyze the images in full resolution more often tend to give more accurate answers, but this was not observed. There are two likely explanations for this: by opening the image in full resolution the user might be able to perceive more details in the image, and be prone to over-analyzing; and secondly, there might be something suspicious in the image, such as a photographic artifact or a contextual cue, even though it is a T image, influencing the users to open it in full resolution, and then answering it is F. This is corroborated by Table~\ref{tab:corrAllImgFeatures}, which shows that opening an image in Full Resolution is associated with lower confidence, higher Hint use, and longer Time observing the image after asking for a Hint.

The Variance is positively correlated with the T:T class, but also according to Table~\ref{tab:corrAllImgFeatures}, it is associated with lower Confidence, higher Hint use, longer Time observing the image after asking for a Hint, and also opening the image in Full Resolution. Our hypothesis is that an image with higher variance is harder to inspect visually, as it contains more details. Users might default to a T answer after not finding anything suspicious. 

\begin{figure*}[ht]
\centering\subfloat[True image with low Variance.]{\includegraphics[width=1.5in]{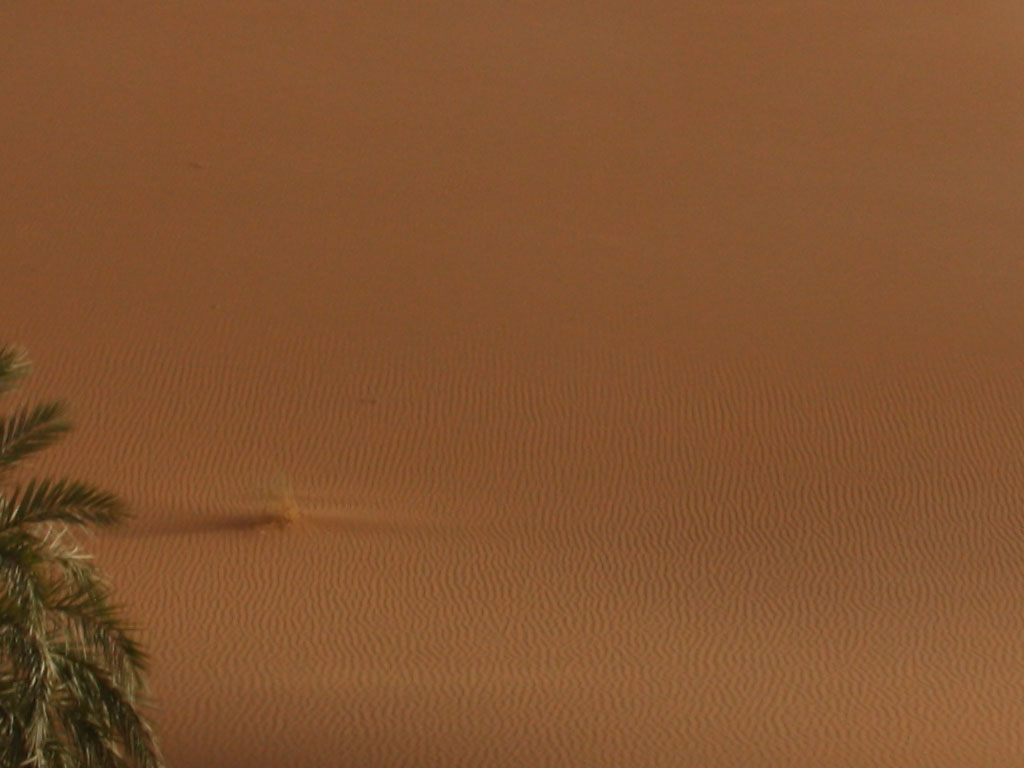}
\label{fig:varExample1_}}
\hfil
\subfloat[Fake image with low Variance.]{\includegraphics[width=1.5in]{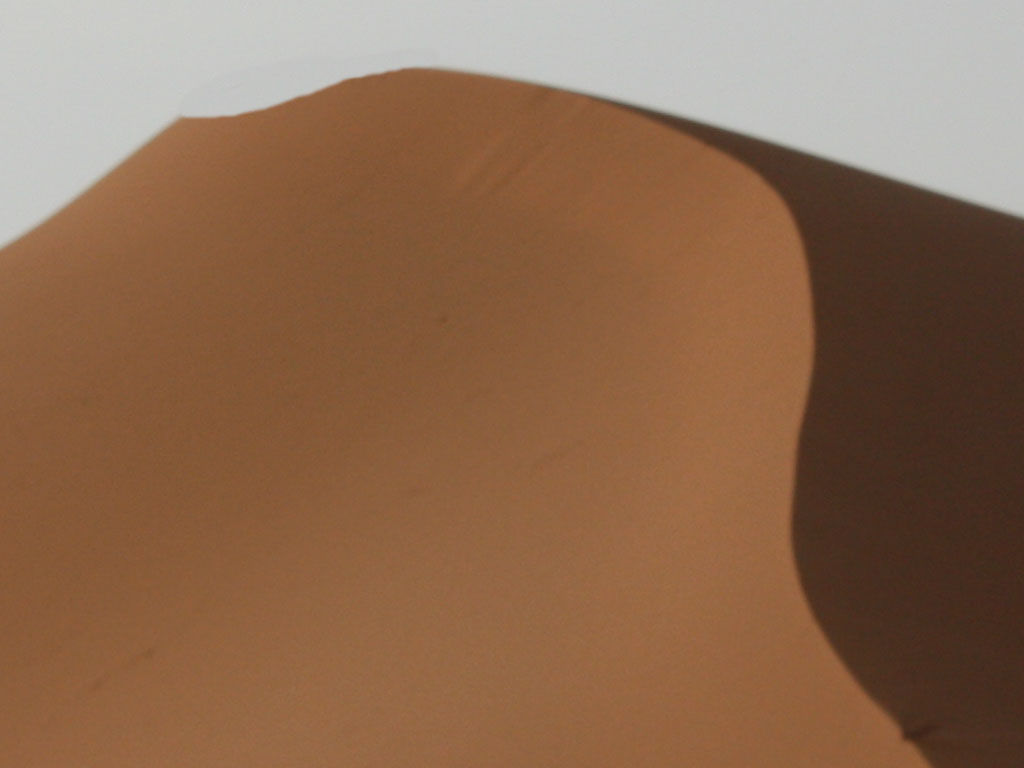}
\label{fig:varExample2_}}
\hfil
\subfloat[True image with high Variance.]{\includegraphics[width=1.5in]{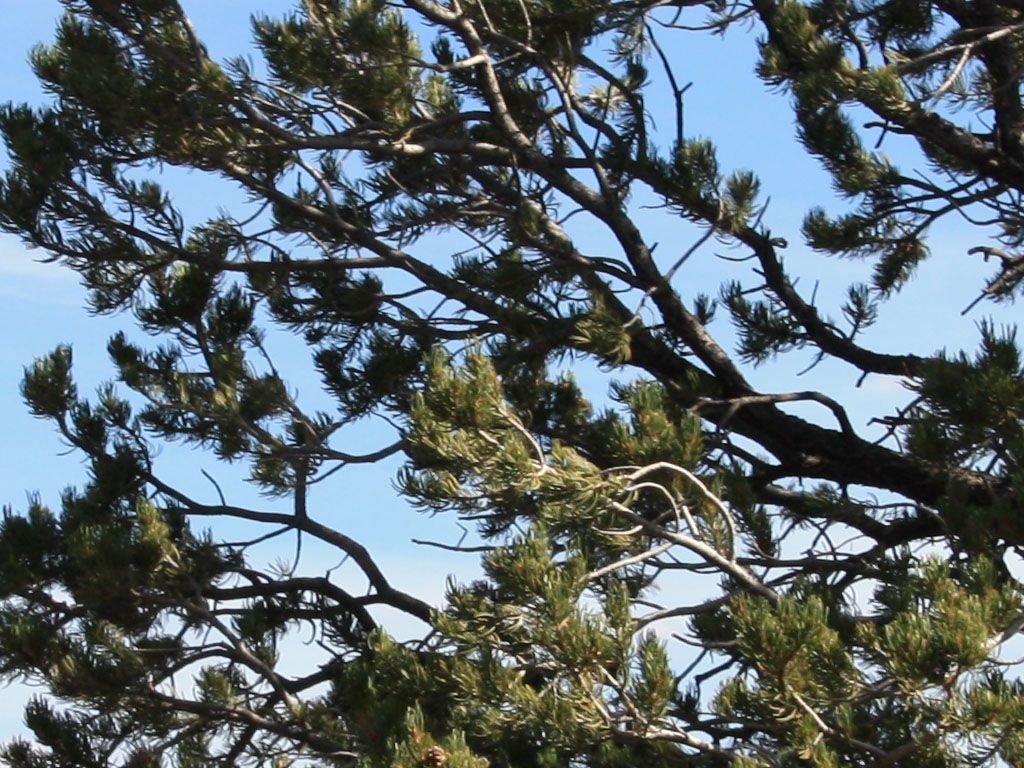}
\label{fig:varExample3_}}
\hfil
\subfloat[Fake image with high Variance.]{\includegraphics[width=1.5in]{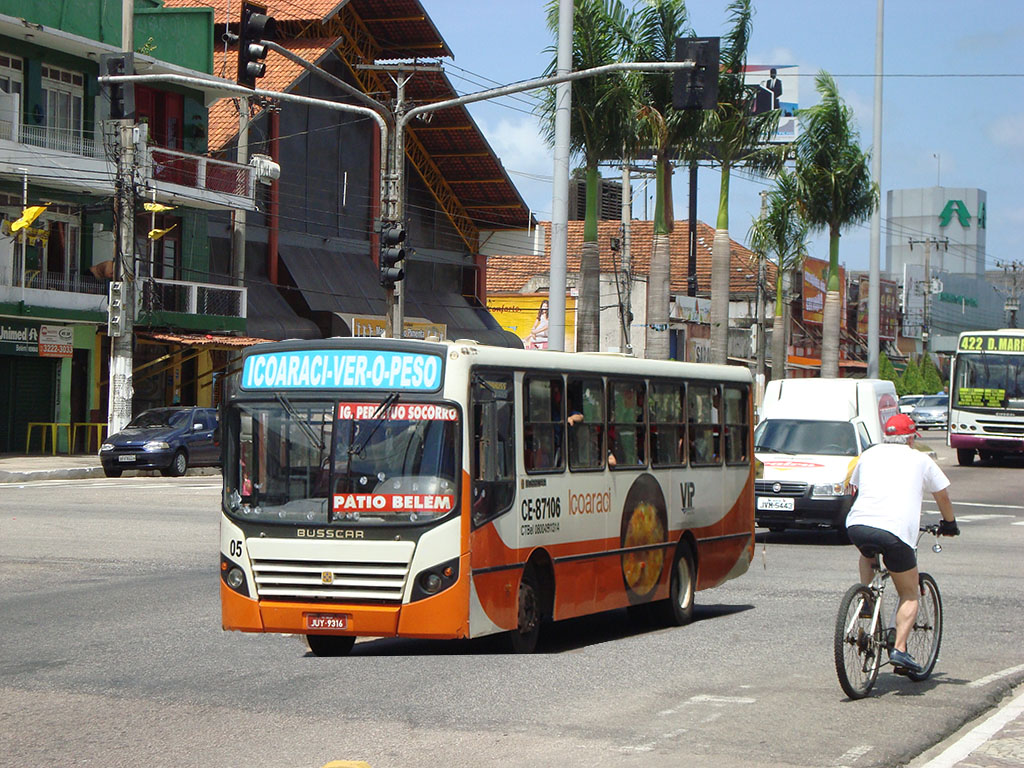}
\label{fig:varExample4_}}
\\
\subfloat[Heatmap of clicks over true image with low Variance.]{\includegraphics[width=1.5in]{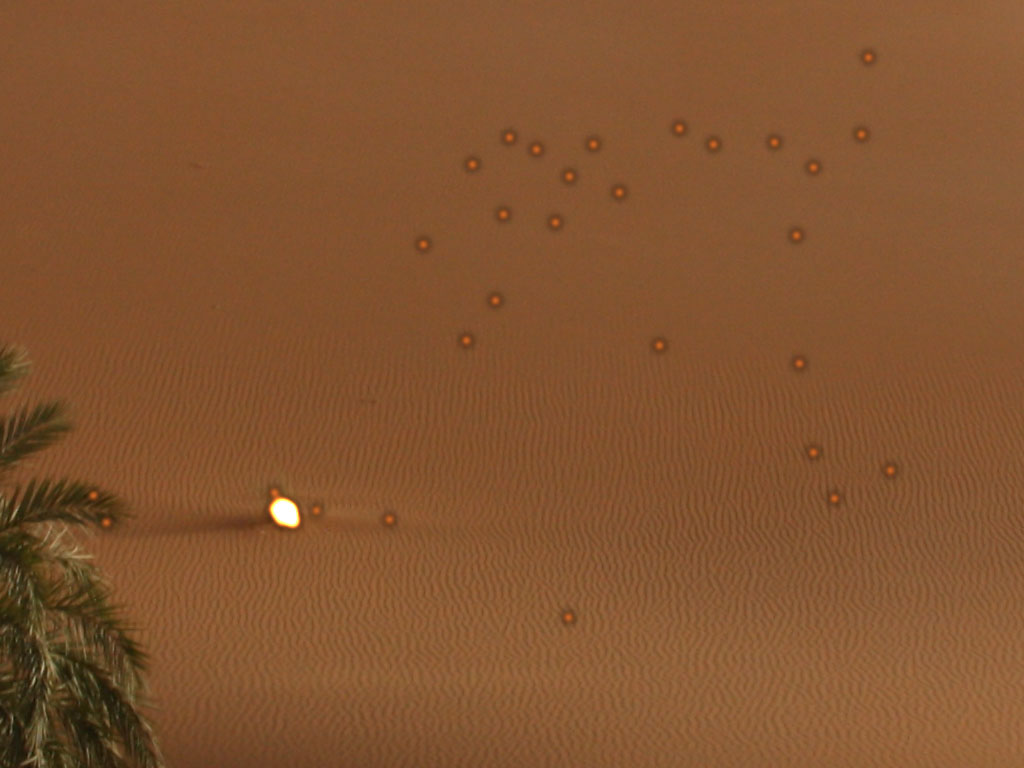}
\label{fig:varExample1}}
\hfil
\subfloat[Heatmap of clicks, and modified area highlighted over fake image with low Variance.]{\includegraphics[width=1.5in]{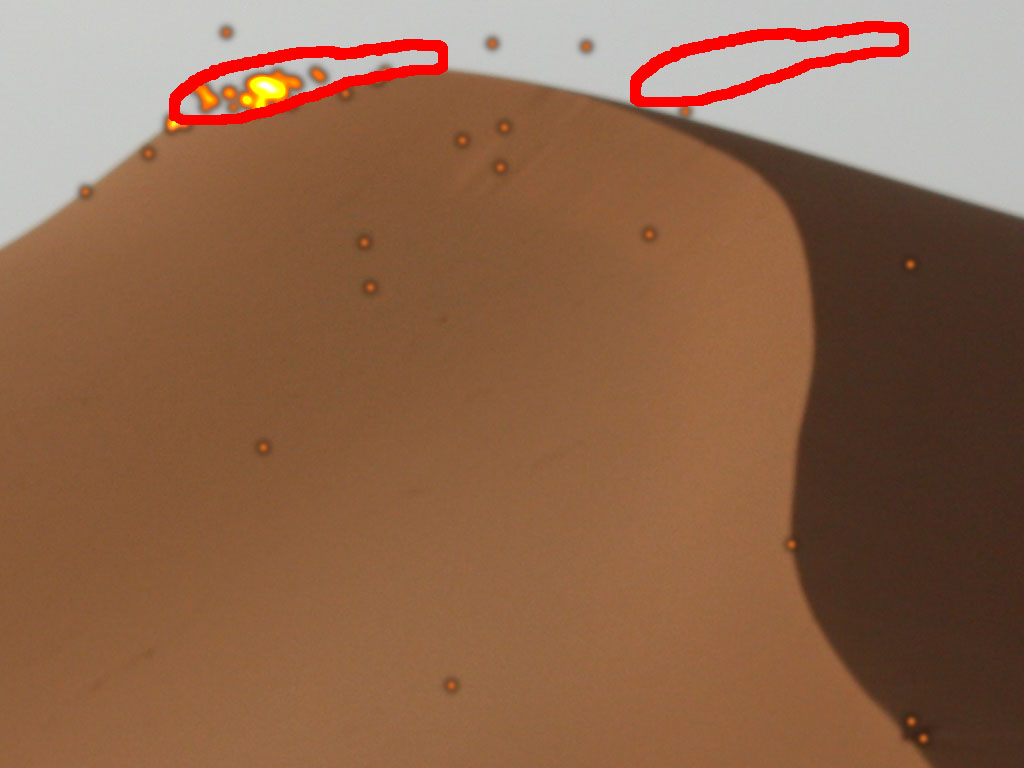}
\label{fig:varExample2}}
\hfil
\subfloat[Heatmap of clicks over true image with high Variance.]{\includegraphics[width=1.5in]{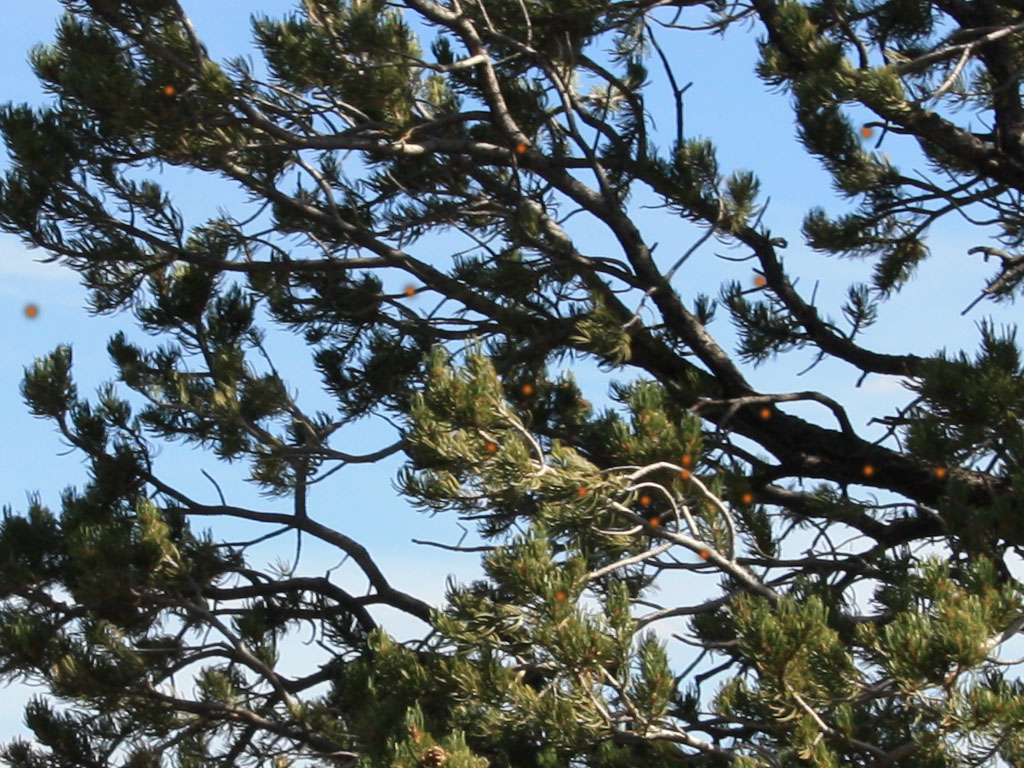}
\label{fig:varExample3}}
\hfil
\subfloat[Heatmap of clicks, and modified area highlighted over fake image with high Variance.]{\includegraphics[width=1.5in]{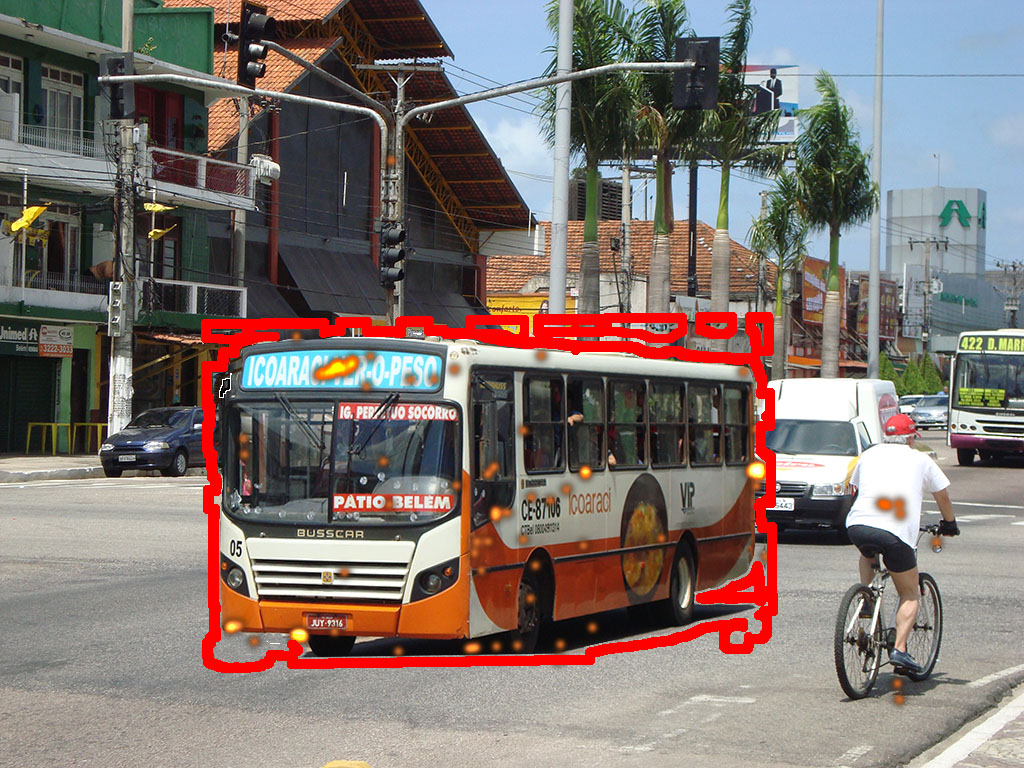}
\label{fig:varExample4}}
\caption{True and fake images with low and high Variance values, respectively. The orange spots indicate the heatmap of user clicks on the image, while the red borders on the fake images outline the edited area. Here the images are shown side-by-side with their overlayed versions to allow the perception of details. }
\label{fig:varExample}
\end{figure*}

Figure~\ref{fig:varExample} displays four examples of images that appear on the test, both T and F with high and low variance. The overlayed versions on the bottom also outline the edited areas and show what parts of the image users clicked to provide evidence of forgery. In Figure~\ref{fig:varExample1}, a T image with low Variance, the users promptly suspected of the smoothed sand patterns and the dune being moved by the wind, and it received 50 F answers. On the other hand, Figure~\ref{fig:varExample3}, which is also T but with high Variance, had an average observation time 6 seconds longer and only 16 F answers.

Observing the clicking patterns, which are represented by heat maps in the overlayed images, can also provide interesting insight on the users' decision process. The majority of clicks on Figure~\ref{fig:varExample1_} is concentrated on a wave of sand provoked by the wind, as it is the most visually striking element of the image. In Figure~\ref{fig:varExample4_}, some users noticed residues from the splicing composition, specially on the back of the bus, and used it as evidence. The shadows of the bus and the biker, which is not forged, caught the attention of several users that clicked on it. Finally, the signs and writing on the bus, most notably the name of the line, were used as evidence of forgery. It is possible that users suspected the signs were altered, and not the bus itself that was spliced, but since we cannot determine their exact reasoning any click on the bus must be considered valid evidence.

\begin{table*}[t]
\centering
\begin{tabular}{|l|c c|c c|c c|}
\hline
& F:Fv $\rho$ & F:Fv $p$ & F:Fi $\rho$ & F:Fi $p$ & F:T $\rho$ & F:T $p$\\
\hline
Confidence &\ \color{Blue}{0.785}&\color{Blue}{1e-50}&  \color{BrickRed}{-0.201} &\color{BrickRed}{0.048}  &\color{BrickRed}{-0.799}&\color{BrickRed}{1e-08}\\
Time       &                    -0.196 & 0.054                   &\ \color{Blue}{0.209}&\color{Blue}{0.040}&\ 0.124                 & 0.225 \\
Time after Hint  & \color{BrickRed}{-0.483}&\color{BrickRed}{5e-07}&\ \color{Blue}{0.201}&\color{Blue}{0.049}&\ \color{Blue}{0.456}&\color{Blue}{2e-06} \\
Hint Proportion      & \color{BrickRed}{-0.552} & \color{BrickRed}{4e-09} &\ \color{Blue}{0.286}&\color{Blue}{0.004}&\ \color{Blue}{0.493}&\color{Blue}{2e-07}\\
Full Resolution         & -0.149 & 0.146 &\ 0.165 & 0.107 &\ 0.091 & 0.374 \\
Edited Area  &\ 0.060 & 0.563 & -0.151 & 0.137 &\ 0.004 & 0.965 \\
Luminance  & -0.271 & 0.790 & -0.598 & 0.558 &\ 0.095 & 0.351 \\
Variance   &\ 0.019 & 0.850 & -0.110 & 0.279 &\ 0.104 & 0.306 \\
\hline
\end{tabular}
\caption{Correlation and respective $p$-value of image features for fake images and the F:Fv,F:Fi and F:T answer classes. Here, blue denotes positive correlation with acceptable $p$-value ($p<0.05$), and red denotes negative correlation with acceptable $p$-value. Black values do not satisfy the threshold and the null hypothesis cannot be rejected.}
\label{tab:corrFakeImgs}
\end{table*}

When estimating correlations among multiple answer classes simultaneously, it is slightly harder to reject the null hypothesis with the amount of evidence obtained. For F images there are only 97 samples (each F image represents a sample, but over 8,667 answers are used to calculate their features) and three classes: F:Fv, F:Fi and F:T. Furthermore, the fake images are split into three types: erasing, copy-paste and splicing forgeries; all quite different in nature. Nevertheless, it is possible to outline several significant correlations, as can be seen in Table~\ref{tab:corrFakeImgs}.

The most notable correlations in this Table are related to Confidence: when users recognize something suspicious they will be very confident in the answer, but will remain in doubt if no clear evidence can be found. The negative correlation between Confidence and the F:Fi class is probably caused by users actively guessing. Since they are unsure in their response, and are knowingly providing a guess, they use a lower confidence value. Time, Time after Hint, and Hint Proportion all have significant correlations, as users try to obtain additional insight on the image nature. Ultimately, they fail to increase performance. This means that the more challenging a forgery is, users will spend more time inspecting it and asking for hints, but for the truly hard ones nothing actually helps. 

The collected data suggests no correlation between the edited area and accuracy, contrary to our expectations. It is intuitive to think that the larger area the forgery covers on the image, the more likely for it to be spotted. The closest correlation found was to the class F:Fi with $\rho=-0.151$ and $p=0.137>0.05$, which could suggest that smaller edited areas might elude users. However, the null hypothesis could not be rejected in this case.

The correlations between all image features are exposed on Table~\ref{tab:corrAllImgFeatures}, where the same pattern can be observed: Confidence, which is the most straightforward indicator of image difficulty correlates negatively with all other image features, except for Luminance. From the two physical aspects analyzed for all images, Luminance and Variance, the evidence strongly suggests the first one has no direct correlation to neither performance nor answering behavior. Variance, in turn, can be associated with almost all difficulty traits. Its biggest correlation is with the Time Hint, meaning that when users asked for hints, the local variability of colors had a heavy influence on how much time they inspected it afterward.


\begin{table*}[t]
\centering
\begin{tabular}{|l|c c|c c|c c|c c|c c|c c|c c|}
\hline
 & C $\rho$ & C $p$ & T $\rho$ & T $p$ & TH $\rho$ & TH $p$ & H $\rho$ & H $p$ & FR $\rho$ & FR $p$ & L $\rho$ & L $p$ & V $\rho$ & V $p$ \\
\hline
P  &\ \color{Blue}{0.200} & \color{Blue}{0.007} & -0.098 & 0.192 & -0.119 & 0.114 & -0.094 & 0.212 & -0.026 & 0.727 & -0.045 & 0.556 &\ 0.051 & 0.500 \\
C  &-       &-      & \color{BrickRed}{-0.245} & \color{BrickRed}{0.001} & \color{BrickRed}{-0.579} & \color{BrickRed}{3e-17} & \color{BrickRed}{-0.684} & \color{BrickRed}{9e-26} & \color{BrickRed}{-0.221} & \color{BrickRed}{0.003} & -0.001 & 0.992 & \color{BrickRed}{-0.149} & \color{BrickRed}{0.048} \\
T  &-       &-      &-       &-      &\ \color{Blue}{0.384} & \color{Blue}{1e-07} &\ \color{Blue}{0.337} & \color{Blue}{4e-06} &\ 0.037 & 0.627 & -0.030 & 0.691 &\ 0.118 & 0.117 \\
TH &-       &-      &-       &-      &-       &-      &\ \color{Blue}{0.764} & \color{Blue}{3e-35 }&\ \color{Blue}{0.252} & \color{Blue}{0.001} &\ 0.054 & 0.479 &\ \color{Blue}{0.374} & \color{Blue}{2e-07} \\
HP  &-       &-      &-       &-      &-       &-      &-       &-      &\ \color{Blue}{0.205} & \color{Blue}{0.006} &\ 0.073 & 0.335 &\ \color{Blue}{0.169} & \color{Blue}{0.025} \\
FR &-       &-      &-       &-      &-       &-      &-       &-      &-       &-      & -0.094 & 0.212 &\ \color{Blue}{0.177} & \color{Blue}{0.019} \\
L  &-       &-      &-       &-      &-       &-      &-       &-      &-       &-      &-       &-      &\ \color{Blue}{0.154} & \color{Blue}{0.041} \\
\hline
\end{tabular}
\caption{Correlation and respective $p$-value between different image features for all images. Here, blue denotes positive correlation with acceptable $p$-value ($p<0.05$), and red denotes negative correlation with acceptable $p$-value. Black values do not satisfy the threshold and the null hypothesis cannot be rejected.}
\label{tab:corrAllImgFeatures}
\end{table*}

\begin{figure}[ht]
  \centering
  \includegraphics[width=3in]{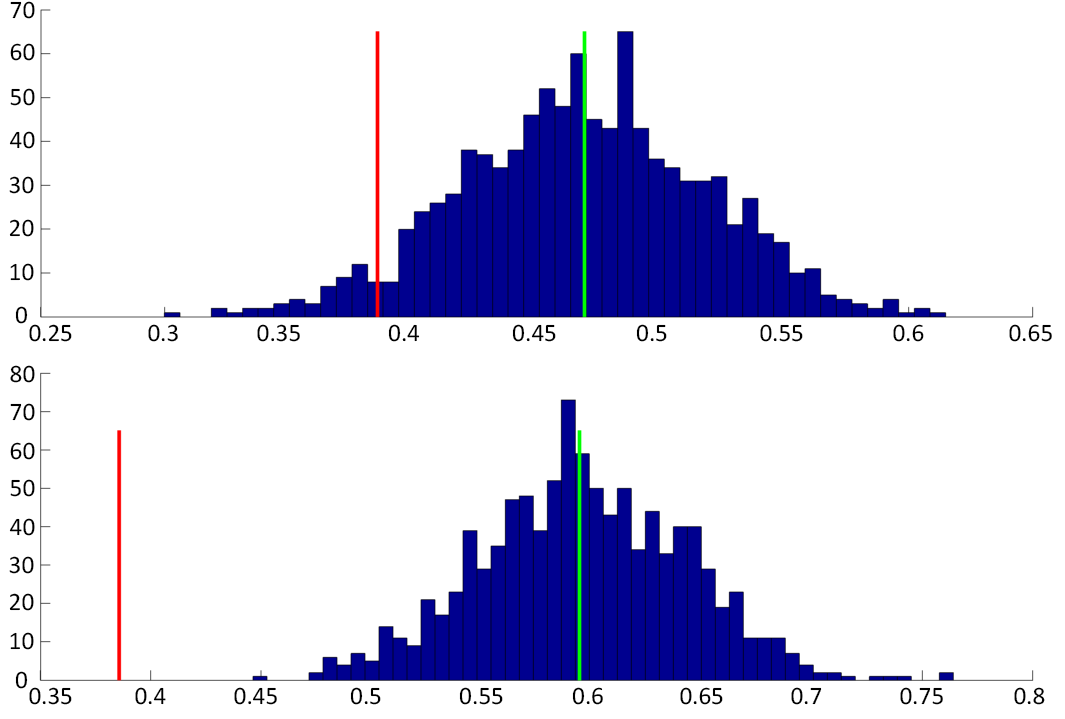}
  \caption{Results of the performance distribution on the resampling process for Copy-Paste (top) and Splicing (bottom) compared to Erasing images. The green line represents the average performance for all images of that type (35 for Copy-Paste and 42 for Splicing), while the red line is the average performance for Erasing forgery.}
	\label{fig:resamplingPerformance}
\end{figure}

The F images consist of 20 Erasing images, 35 Copy-Paste images and 42 Splicing images. The average accuracy for each of those forgery types was 0.385, 0.469, and 0.594, respectively. This suggests that Erasing images are harder to identify than Copy-Paste, which are harder than splicing. Since we have a different number of images from each type, it is necessary to account for any possible bias that this could cause.

To assess this, we re-sampled the splicing and copy-paste images in smaller groups, selecting 20 images at each time\footnote{In other words, the combinations $C_{20}^{35}$ and $C_{20}^{42}$.} (the number of the smaller set, erasing). On total, 1,000 resampling operations for each of the two sets, and calculated the average accuracy for each time. 

The results of the combinatorial process are displayed in Figure~\ref{fig:resamplingPerformance}, where the histogram represents the distribution of average accuracy for each resampling. The green line is the average accuracy considering all images from that type together, and the red line the average accuracy for the Erasing class. It is clear that the Splicing images have superior average accuracy than Erasing images. The results of our study indicate that Erasing images are the most challenging ones for users to identify, and Copy-Paste generally is harder than Splicing.

\subsection{Anecdotal Observations}
\label{sec:Anecdotal}
After performing the study, we analyzed feedback from users to complement the results. This included e-mails, information sent using the feedback form on the website, contact through electronic messaging and presential meetings. Only a small sample of users provided some form of feedback (less than 10\% of participants), and there was no particular methodology for this analysis, so the reliability of these results is limited. Nevertheless, they provide insight on how users interacted with the study.

All questioned users, including those with high accuracy, reported that they found the study hard in general. Users that were questioned presentially were shown images from the test and inquired about their answer and justification. From their responses, it appears that there is no consensus among users on what seems to be an "obvious" forgery. Even the most explicit forgeries have eluded some users, and they have provided justification for their mistakes. Among the provided answers on why they missed any particular image, the most common responses were:

When missing a T image, by saying it was F, or missing an F image by providing wrong evidence:
\begin{itemize}
\item The user was fooled by a photographic artifact, such as lens flare, residue on the lens or even by the image exposure;
\item After asking for a hint and spending a large amount of time analyzing the image, the user felt compelled to guess that something was manipulated;
\item Something in the context of the scene depicted in the image felt wrong.
\end{itemize}

When missing a fake image, by saying it was true:
\begin{itemize}
\item The user did not pay much attention to that particular part of the image, even after asking for a hint.
\item They felt the image was too cluttered or there was too much in the image to be analyzed.
\item They looked at the manipulated region and found it suspect but plausible, or were not expecting it to be a manipulation at all.
\end{itemize}

Some of these items are directly supported by our results, as discussed in this section. 

An important observation is that users rely strongly on context in order to make decisions. Several images depict people in social situations, some of which appear in more than one image. More than one user reported being suspicious of particular characters after they were used in a forgery. Some users went as far as to imply social relations, for instance assuming that two people depicted in the images were a couple and suspecting when one of them appeared together with someone else. Images of cars and traffic also prompted contextual analysis by users. The most notable occurrences were images where a car was spliced driving in the wrong lane and a another in which the "Mercedes Benz" logo was spliced over a Volkswagen car.


%% file: sec/relatedwork.tex
\section{Related Work}
\label{sec:Related}

To the best of our knowledge, ours is the most comprehensive study to evaluate people's ability to identify false images. 
There have been, however, related studies with different scopes and scales.

On the field of perception, \etal{Ostrovsky}~\cite{Ostrovsky2005} explore how different lighting configurations influence user perception. The study explored different forms of visual stimuli: 3D computer-generated scenes, photographs, and pictures were shown to the subjects. The authors measured the response time and accuracy in detecting lighting irregularities in different configurations. The study concluded that while evaluating a small set of objects it is easy to assess outliers, the task becomes very hard in complex scenes with different objects and light interactions. Unlike our study, this one systematically evaluates the perception of a specific feature (i.e., lighting) using a small number of subjects (17), but a strictly controlled environment and experiment.

\etal{Farid}~\cite{FaridHumanVisual} explore how good users are at detecting irregularities in geometry, shades and reflections, discussing their forensic implications. For this purposes, tests are performed with twenty subjects observing different pictures and trying to identify tampering. Their results show that humans are inept at perceiving inconsistencies in shadows, reflections, and planar perspective distortions. Furthemore, forensics solutions are presented that could be used to help users on identifying these types of forgeries. This study differs from ours by focusing on the perception of specific image features and by giving special attention to a small, controlled group.

Another work by \etal{Farid}~\cite{FaridFaces} assessed the subjects' ability to distinguish photographs of human faces from computer generated faces. 
The results show that while humans can reliably distinguish photographs from CG models under various circumstances, modern CG techniques and good 3D modeling pose very hard challenges.

\etal{Carvalho}~\cite{Tiago2013} proposed a technique to detect forgery based on color classification of the scene illuminants. To validate their approach, the authors created an image database with true and false images\footnote{Using splicing forgery.}, and perform a variety of tests, some of which involved human subjects. Similar to our study, theirs was not focused on a particular feature of human perception, and obtained around 2,000 individual answers over 200 images. The reported accuracy of the tested subjects was of 64.7\%, identifying only 38.3\% the false images. 
The authors used a binary classification (true or false images), implying that the accuracy could have been influenced by subject guessing the right answer based on incorrect justification.

All of these works support our findings that humans are not generally good at identifying edits in digital images. 

%% file: sec/conclusion.tex
\section{Conclusion}
\label{sec:Conclusion}


This study analyzed human's ability to detect forgery in digital images. Its findings suggest that humans can be easily fooled. Participants not only performed poorly at identifying forgery in images, but they often doubted the authenticity of pristine pictures. It was also demonstrated that the nature of an image and its features may affect ones ability to detect forgeries. As such, further work is required to better understand the relevant aspects involved in such observed behavior.

\subsection{Main Findings}

The core finding of this study is that users have a poor performance in identifying modifications in images, guessing right only around 58\% of the time, and only identifying 46.5\% of actual forgeries. Aside from age, the background had little to no effect on a user's ability to identify forgery. The behavior (time, hints, confidence, etc.) while answering the study had the biggest impact on the success rate. This leads us to believe that there might be strategies and good practices to aid users in spotting modified images. This should be verified in further studies.

These findings are of special importance to the forensics community, for two main reasons. They show the importance of having tools to help us to authenticate and investigate images. Secondly, it is common for forensics methods to require user input, either by selecting key points, an area in the image, or tweaking parameters. Human's low ability to suspect of image forgeries directly impacts the performance of such techniques. 

The results of our study indicate that erasing is the hardest type of forgery to detect, followed by copy-paste, and splicing. This is interesting for the forensics community, because it highlights which types of forgery are more challenging and relevant. 

\subsection{Limitations and Future Work}


The main limitation of our paper is the small amount of copy-paste and erasing images used in the study, compared to splicing. This was caused mainly due to the lack of such images in public forensics databases. To solve this issue, calculated the statistics using re-sampled sets of data (Section~\ref{sec:Images}), which is not ideal. 

Given the nature of the study and the amount of uncovered data, several things can be further investigated. For example, there are 8,160 points of evidence provided over the 177 images in the form of user clicks. This data could provide insights on what kinds of objects or image elements the subjects are more prone to suspect.

It is also possible to test the influence of different image-composition techniques on the quality of the forgery. A \textit{splicing} forgery, for instance, can be done by simply cutting and pasting a region from an image into another, or by using sophisticated tools. Alpha Matting~\cite{Gastal2010SharedMatting}, and gradient domain composition techniques~\cite{Farbman2009}~\cite{Sunkavalli2010}~\cite{Darabi12} are able to blend two images, creating visually imperceptible compositions. They are not perfect, however, and differences in perspective or illumination may tip users off. A study using our methodology focused on different types of \textit{splicing} forgery could help us better identify dangerous composition techniques.

\section*{Acknowledgements}

This work was supported by funding from CAPES and CNPq-Brazil (Grants 306196/2014-0, 482271/2012-4, and 311862/2012-8). Special thanks to Victor Adriel, for helping with user interaction issues, and Francisco Borja and Gerson Groth for assistance on the early implementation of the webservice. The online test was hosted on the PPGC UFRGS\footnote{http://ppgc.inf.ufrgs.br/} infrastructure. 


%% file: sec/appendix2.tex
\appendices

\section{Influence of User Level on Accuracy}
\label{sec:influenceLevel}

Due to the extensive size of our study, it is hard to guarantee that users will fully complete it. As was shown in Section~\ref{sec:UserBackground}, only 24 users answered all images, and there is a large variance in the amount of answers per user. For this reason, it is important to determine if a users' performance on the test changes as he answers more images.

\begin{figure}[ht]
  \centering
  \includegraphics[width=3in]{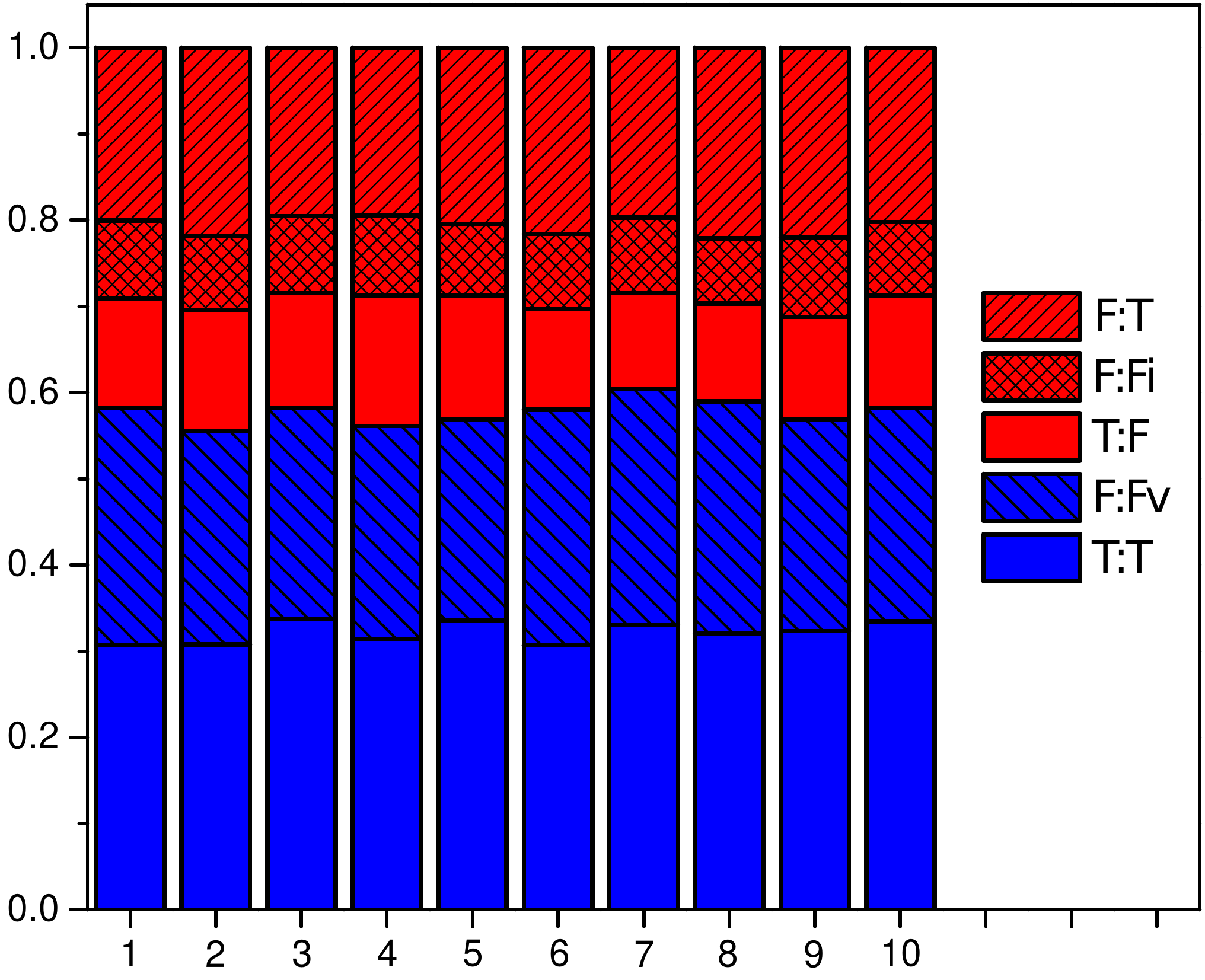}
  \caption{Distribution of answers classes per user level. This graph uses the current level of the user at the time he provided a particular answer.}
	\label{fig:percentLVL}
\end{figure}

Figure~\ref{fig:percentLVL} shows the distribution of answer classes for all different user levels. The bottom two classes represent the T:T and F:Fv classes, respectively, and their stacked value corresponds to the overall accuracy of that user level. The user level, in this case, is logged at the moment of the answer. This means that users will have answers being accounted on different groups if they advanced levels. A user that completed the test will have answers spread across all groups.

This data shows that there is no significant change in the accuracy of users when they progress on the test, and the average accuracy is slightly below 60\% for all levels. This is good for our methodology, because it allows to treat all answers equally regardless of the user's progress in the test.

\begin{figure}[ht]
  \centering
  \includegraphics[width=3in]{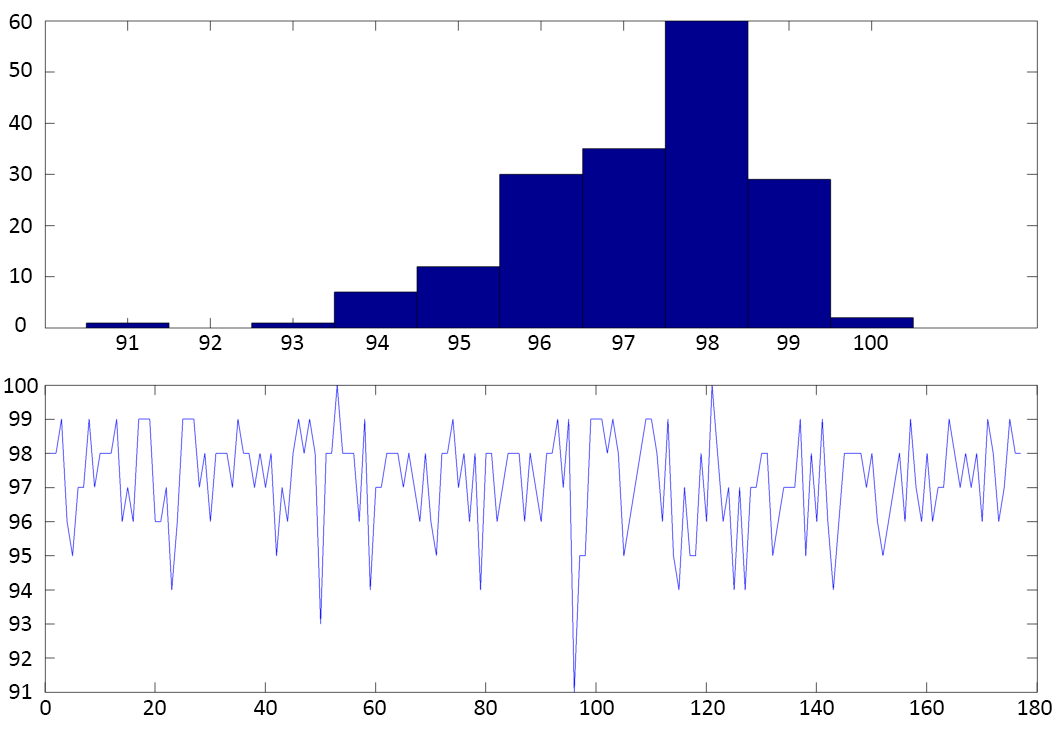}
  \caption{Amount of answers received for each image. On top, the histogram of answers for all images, and on bottom the graph of answers received for each image.}
	\label{fig:answersPerImg}
\end{figure}

The algorithm for randomizing the image order was responsible for reducing the influence of user level. Every time a user opened the test page, or answered an image and requested a new one, the 20 least answered images were determined. One of these 20 least answered images was then selected, at random. Since several users were online providing answers at the same time, the least answered images constantly changed. This was also important to guarantee all images had a similar number of answers, which is ideal for comparing data. The majority of the 177 images received between 96 and 99 user answers, as can be seen in Figure~\ref{fig:answersPerImg}.

%% file: bare_jrnl.bbl
\begin{thebibliography}{10}
\providecommand{\url}[1]{#1}
\csname url@samestyle\endcsname
\providecommand{\newblock}{\relax}
\providecommand{\bibinfo}[2]{#2}
\providecommand{\BIBentrySTDinterwordspacing}{\spaceskip=0pt\relax}
\providecommand{\BIBentryALTinterwordstretchfactor}{4}
\providecommand{\BIBentryALTinterwordspacing}{\spaceskip=\fontdimen2\font plus
\BIBentryALTinterwordstretchfactor\fontdimen3\font minus
  \fontdimen4\font\relax}
\providecommand{\BIBforeignlanguage}[2]{{%
\expandafter\ifx\csname l@#1\endcsname\relax
\typeout{** WARNING: IEEEtran.bst: No hyphenation pattern has been}%
\typeout{** loaded for the language `#1'. Using the pattern for}%
\typeout{** the default language instead.}%
\else
\language=\csname l@#1\endcsname
\fi
#2}}
\providecommand{\BIBdecl}{\relax}
\BIBdecl

\bibitem{SurveyPiva}
A.~Piva, ``An overview on image forensics,'' \emph{ISRN Signal Processing},
  vol. 2013, pp. 1--22, 2013.

\bibitem{Rocha_2011}
\BIBentryALTinterwordspacing
A.~Rocha, W.~Scheirer, T.~Boult, and S.~Goldenstein, ``Vision of the unseen:
  Current trends and challenges in digital image and video forensics,''
  \emph{ACM Comput. Surv.}, vol.~43, no.~4, pp. 26:1--26:42, Oct. 2011.
  [Online]. Available: \url{http://doi.acm.org/10.1145/1978802.1978805}
\BIBentrySTDinterwordspacing

\bibitem{FontaniDataFusion}
M.~Fontani, A.~Bonchi, A.~Piva, and M.~Barni, ``Countering anti-forensics by
  means of data fusion,'' in \emph{Proc. SPIE 9028, Media Watermarking,
  Security, and Forensics 2014}, vol. 9028, 2014, pp. 90\,280Z--90\,280Z--15.

\bibitem{Tiago2013}
T.~J. de~Carvalho, C.~Riess, E.~Angelopoulou, H.~Pedrini, and
  A.~de~Rezende~Rocha, ``Exposing digital image forgeries by illumination color
  classification,'' \emph{IEEE Transactions on Information Forensics and
  Security}, pp. 1182--1194, 2013.

\bibitem{Cozzolino2014}
D.~Cozzolino, G.~Poggi, and L.~Verdoliva, ``Copy-move forgery detection based
  on patchmatch,'' in \emph{IEEE International Conference on Image Processing},
  2014.

\bibitem{Chierchia2014}
G.~Chierchia, G.~Poggi, C.~Sansone, and L.~Verdoliva, ``A bayesian-mrf approach
  for prnu-based image forgery detection,'' \emph{Information Forensics and
  Security, IEEE Transactions on}, vol.~PP, no.~99, pp. 1--1, 2014.

\bibitem{Chierchia2013}
------, ``Prnu-based forgery detection with regularity constraints and global
  optimization,'' in \emph{Multimedia Signal Processing (MMSP), 2013 IEEE 15th
  International Workshop on}, 2013, pp. 236--241.

\bibitem{Bianchi2012}
T.~Bianchi and A.~Piva, ``Image forgery localization via block-grained analysis
  of jpeg artifacts,'' \emph{IEEE Transactions on Information Forensics and
  Security}, vol.~7, no.~3, pp. 1003 -- 1017, 2012.

\bibitem{Qu2014}
Z.~Qu, W.~Luo, and J.~Huang, ``A framework for identifying shifted double jpeg
  compression artifacts with application to non-intrusive digital image
  forensics,'' \emph{Science China Information Sciences}, vol.~57, no.~2, pp.
  1--18, 2014.

\bibitem{ritterfeld2009}
\BIBentryALTinterwordspacing
U.~Ritterfeld, M.~Cody, and P.~Vorderer, \emph{Serious Games: Mechanisms and
  Effects}.\hskip 1em plus 0.5em minus 0.4em\relax Routledge, 2009. [Online].
  Available: \url{http://books.google.com.br/books?id=GwPf7tbO5mgC}
\BIBentrySTDinterwordspacing

\bibitem{trivedi2002probability}
K.~Trivedi, \emph{Probability and statistics with reliability, queuing, and
  computer science applications}.\hskip 1em plus 0.5em minus 0.4em\relax New
  York: Wiley, 2002.

\bibitem{luminance1996}
\BIBentryALTinterwordspacing
M.~Stokes, M.~Anderson, S.~C. Anderson, S.~Chandrasekar, and R.~Motta. (1996,
  November) A standard default color space for the internet - srgb. [Online].
  Available: \url{http://www.w3.org/Graphics/Color/sRGB}
\BIBentrySTDinterwordspacing

\bibitem{Ostrovsky2005}
P.~S. Y.~Ostrovsky, P.~Cavanagh, ``Perceiving illumination inconsistencies in
  scenes,'' \emph{Perception}, vol.~34, no.~11, pp. 1301--1314, 2005.

\bibitem{FaridHumanVisual}
H.~Farid and M.~Bravo, ``Image forensic analyses that elude the human visual
  system,'' \emph{SPIE Symposium on Electronic Imaging}, 2010.

\bibitem{FaridFaces}
------, ``Perceptual discrimination of computer generated and photographic
  faces,'' \emph{Digital Investigation}, 2012.

\bibitem{Gastal2010SharedMatting}
E.~S.~L. Gastal and M.~M. Oliveira, ``Shared sampling for real-time alpha
  matting,'' \emph{Computer Graphics Forum}, vol.~29, no.~2, pp. 575--584, May
  2010, proceedings of Eurographics.

\bibitem{Farbman2009}
\BIBentryALTinterwordspacing
Z.~Farbman, G.~Hoffer, Y.~Lipman, D.~Cohen-Or, and D.~Lischinski, ``Coordinates
  for instant image cloning,'' \emph{ACM Trans. Graph.}, vol.~28, no.~3, pp.
  67:1--67:9, Jul. 2009. [Online]. Available:
  \url{http://doi.acm.org/10.1145/1531326.1531373}
\BIBentrySTDinterwordspacing

\bibitem{Sunkavalli2010}
K.~Sunkavalli, M.~K. Johnson, W.~Matusik, and H.~Pfister, ``Multi-scale image
  harmonization,'' \emph{ACM Transactions on Graphics}, vol.~29, no.~4, pp.
  125:1--125:10, 2010.

\bibitem{Darabi12}
S.~Darabi, E.~Shechtman, C.~Barnes, D.~B. Goldman, and P.~Sen, ``{I}mage
  {M}elding: Combining inconsistent images using patch-based synthesis,''
  \emph{ACM Transactions on Graphics (TOG) (Proceedings of SIGGRAPH 2012)},
  vol.~31, no.~4, pp. 82:1--82:10, 2012.

\end{thebibliography}
